\documentclass[fleqn,usenatbib]{mnras}

\usepackage{newtxtext,newtxmath}

\usepackage[T1]{fontenc}

\DeclareRobustCommand{\VAN}[3]{#2}
\let\VANthebibliography\thebibliography
\def\thebibliography{\DeclareRobustCommand{\VAN}[3]{##3}\VANthebibliography}

\usepackage{graphicx}	
\usepackage{amsmath}	

\newcommand{\thC}{\ensuremath{\theta_{C}}}
\newcommand{\epse}{\ensuremath{\epsilon_{e}}}
\newcommand{\epsB}{\ensuremath{\epsilon_{B}}}
\newcommand{\xiN}{\ensuremath{\xi_{N}}}
\newcommand{\xcen}{\ensuremath{\tilde{x}_{\mathrm{c}}}}
\newcommand{\afterglowpy}{{\texttt{afterglowpy}}}

\title[Scaling independent of jet structure and dynamics]{Scaling relations for gamma-ray burst afterglow light curves and centroid motion independent of jet structure and dynamics}

\author[Van Eerten \& Ryan]{
Hendrik J. van Eerten,$^{1}$\thanks{E-mail: hjve20@bath.ac.uk}
Geoffrey S. Ryan,$^{2}$
\\
$^{1}$Physics Department, University of Bath, Claverton Down, Bath BA2 7AY, UK\\
$^{2}$Perimeter Institute for Theoretical Physics, Waterloo, Ontario N2L 2Y5, Canada\\
}

\date{Accepted XXX. Received YYY; in original form ZZZ}

\pubyear{2023}

\begin{document}
\label{firstpage}
\pagerange{\pageref{firstpage}--\pageref{lastpage}}
\maketitle

\begin{abstract}
Models for gamma-ray burst afterglow dynamics and synchrotron spectra are known to exhibit various scale invariances, owing to the scale-free nature of fluid dynamics and the power-law shape of synchrotron spectra. Since GRB 170817A, off-axis jet models including a lateral energy structure in the initial outflow geometry have gained in prominence. Here we demonstrate how the scale-invariance for arbitrary jet structure and dynamical stage can be expressed locally as a function of jet temporal light curve slope. We provide afterglow flux expressions and demonstrate their use to quickly assess the physical implications of observations. We apply the scaling expressions to the Swift XRT sample, which shows a spread in observed fluxes, binned by light curve slope at time of observation, that increases with increasing light curve slope. According to the scaling relations, this pattern is inconsistent with a large spread in environment densities if these were the dominant factor determining the variability of light curves. We further show how the late Deep Newtonian afterglow stage remains scale-invariant but adds distinct spectral scaling regimes. Finally, we show that for given jet structure a universal curve can be constructed of the centroid offset, image size and ellipticity (that can be measured using very-large baseline interferometry) versus observer angle, in a manner independent of explosion energy and circumburst density. Our results apply to any synchrotron transient characterized by a release of energy in an external medium, including supernova remnants, kilonova afterglows and soft gamma-repeater flares.
\end{abstract}




\section{Introduction}

Gamma-ray bursts (GRB) are sudden gamma-ray transients triggered by the cataclysmic collapse of a massive star or the merger of two neutron stars. The occurrence of a long-lived broadband afterglow has been a key prediction of the cosmological fireball model for GRBs \citep{ReesMeszaros:1992, MeszarosRees:1993}. More generally, any mechanism by which a large amount of energy (about $10^{48-52}$ erg for GRBs) is suddenly released into the circumburst environment will eventually lead to a system of shocks and the emission of predominantly synchrotron radiation from shock-accelerated electrons across a broad spectral range from radio to X-rays, regardless of the prompt emission mechanism. Even though many open questions remain regarding outflow dynamics, launch mechanism and details of the emission, these afterglow counterparts to the GRB prompt emission can be modelled in a relatively straightforward and general manner by coupling a parametrisation of electron shock-acceleration and emission to a dynamical model for the plasma flow \citep{Wijers:1997, Sari:1998}.

GRB afterglow blast waves are known to be highly relativistic, at least initially, and as such their modelling has often relied on the self-similar solution for a relativistic point explosion first presented in \cite{Blandford:1976}. At early times this model not only applies to spherical explosions but also to the radial flow of plasma truncated at some opening angle (the \emph{jetted} outflow associated with afterglows) as long as no causal contact across the shock front is established to induce substantial deformation from radial flow. Being hot and over-pressurized relative to their surroundings, GRB afterglow jets will eventually spread laterally and become trans-relativistic, at which point the self-similar assumption no longer holds. Eventually the jets inevitably become (quasi-)spherical and their dynamics can once again be modelled self-similarly using the well-known Sedov-Taylor-Von Neumann solution \citep{Sedov:1959}. 

The self-similarity of the early and late stage dynamics simplifies the modeling of afterglows, and flux expressions for the broadband synchrotron spectrum have been formulated based on these solutions to great practical effect \citep{Frail:2000, GranotSari:2002}. This includes the formulation of \emph{closure relations} for broadband flux of the form $F_\nu \propto t^\alpha \nu^\beta$ that link the slope $\beta$ associated with a given synchtrotron spectral regime to light curve slope $\alpha$ of a given dynamical regime, typically by solving for a shared variable $p$ that describes the slope of the number density distribution $n_e$ of shock-accelerated electrons as a function of electron Lorentz factor $\gamma_e$, i.e., $n_e \propto \gamma_e^{-p}$ (for an extensive survey of closure relations, see e.g. \citealt{Gao:2013}).

The original self-similar flow closure relations do not apply to the dynamical stage of jet spreading and rely on observers viewing the jet on-axis (and at early times not being able to tell the outflow is jetted rather than spherical, given that GRBs start out as strongly beamed point sources, \citealt{Rhoads:1997}). Including off-axis observations and predictions for the light curve once jet spreading begins to impact its slope instead relies on assumptions for jet spreading dynamics \citep{Rhoads:1999, Sari:1999}. Numerically resolved high-resolution simulations of the long-term evolution of afterglow jets have since confirmed that the spreading behaviour of jets is highly non-linear and does not conform easily to simplified analytical models \citep{ZhangMacFadyen:2009, vanEerten:2011, Wygoda:2011, vanEerten:2012observationalimplications} for the range of jet initial opening angles typically inferred for GRBs\footnote{Extremely narrow jets with an opening angle well below 3 degrees can actually be shown to spread laterally in the exponential regime described by \citealt{Rhoads:1999}. For discussion, see \cite{vanEerten:2018}.}, which limits the applicability of jet-stage closure relations and flux expressions. 

A further complication is that GRB 170817A has firmly established (e.g. \citealt{Troja:2017, Troja:2018, Troja:2019, Lazzati:2018, Lyman:2018, Alexander:2018, Fong:2019, Ghirlanda:2019, TakahashiIoka:2020, Beniamini:2020, Urrutia:2021, Nativi:2022, Garcia-Cifuentes:2024}) that the initial jetted release of energy of afterglow jets is not homogeneous across angles (a \emph{top-hat} jet energy profile), but rather follows that of structured jet models \citep{Meszaros:1998, Rossi:2002}. Closure relations for the rising stage of a structured jet light curve can be formulated \citep{Ryan:2020}, as well as for the decaying slope of a structured jet in the absence of spreading once additional material outside the jet core gradually comes in to view (e.g. \citealt{Beniamini:2022}), but dynamical jet simulations using special relativistic hydrodynamics (SRHD) are required to capture the full light curve from afterglow jets.

However, useful scaling relations remain applicable outside of the self-similar limit even to jet flow computed using SRHD simulations. As shown in \cite{vanEerten:2012boxfit}, practical use (for computing the flux) can be made from rescaling the fluid dynamics conservation laws using dimensional analysis for initial parameters $E_0$ (isotropic-equivalent jet energy along the jet tip) and $\rho_0$ (setting the scale of the circumburst mass density; this can be generalized from a homogeneous medium to a density decreasing with distance as a power-law in radius). Results for different dimensionless jet initial conditions such as the opening angle $\theta_0$ of a top-hat jet, or a range of parametrizations of initial jet structures in the case of non top-hat jets, can be tabulated. Additional model parameters carrying dimension, such as ejecta mass, can be tabulated too, as long as the tabulated values are scaled as dictated by dimensional analysis when applied. This combination of tabulation and rescaling of fluid simulations was first implemented in \texttt{boxfit} \citep{vanEerten:2012boxfit} which linked simulations to a radiative transfer module for computing GRB afterglows.

The scalings can be generalized further to include the characteristic features of synchrotron power-law spectra, in particular their break frequencies $\nu_a$ (synchrotron self-absorption), $\nu_m$ (due to the lower Lorentz factor cut-off $\gamma_m$ of the non-thermal electron population), $\nu_c$ (electron cooling) and peak flux level $F_\textrm{peak}$ (see \citealt{vanEerten:2012scale-invariance} for a first demonstration and \citealt{vanEertenMacFadyen:2013} and \citealt{Granot:2012} for some subsequent discussion). The scaling relations can be used as a basis for software for fitting SHRD simulations to broadband data working directly from spectral templates. A first example of this approach is \texttt{scalefit} (for applications, see e.g. \citealt{Ryan:2015, Zhang:2015, Aksulu:2022, deWet:2023}). \citealt{Lipunov:2017} show how the scalings can be used to rescale data directly.

The discovery of off-axis event GRB 170817A was a further catalyst for GRB analysis using rescaling of simulation output. While \cite{Lazzati:2018} performed an early actual simulation-based fit by rescaling the synchrotron parameters for a jet travelling through non-relativistic ejecta from a neutron star merger (leaving the dynamical scale-invariance unexploited), subsequent works (in particular,  \citealt{WuMacFadyen:2018, WuMacFadyen:2019, Hajela:2019, Gill:2019, McDowellMacFadyen:2023}) modelled the structure of the jet from GRB 170817A and others by dynamical rescaling of simulations that evolve from initial conditions that (except for \citealt{Gill:2019}) did not  describe a top-hat jet.

In this work, we present the afterglow scale invariances for arbitrary jet geometries and circumburst density profiles by showing how the flux is proportional to afterglow model parameters for any light curve slope $\alpha$. This can be used to infer the sensitivity to a given underlying physical model parameter under the assumption that the light curve approximately exhibits power law behaviour around the time of observation. Once a jet structure and environment profile are assumed, the flux equations can be calibrated using e.g. \texttt{afterglowpy} (\citealt{Ryan:2020}, which is based on a structured jet shell model to compute the afterglow). A direct application of this method to the Swift XRT sample is included, demonstrating how the scale-invariance in the presented form can be utilized to extract physical information about jet properties directly from observational data, without requiring the intermediate step of fitting light curves or spectral templates from a simulation or semi-analytical model (which would be limited to a prescribed jet structure used in the simulation or model).

In rare cases, for nearby extremely bright sources such as GRB 030329 \citep{Taylor:2004}, potentially\footnote{This has now been reported by \cite{Giarratana:2023}, while this paper was undergoing peer review.} GRB 221009A \citep{Malesani:2023, OConnor:2023, Williams:2023, Fulton:2023} and very nearby sources such as GRB counterparts to a gravitational wave (GW) detection, very-large-baseline-interferometry (VLBI) may be possible \citep{Mooley:2018, Ghirlanda:2019, Mooley:2022}. We show that for given jet structure a universal curve can be constructed of the centroid offset, images size and elipticity versus observer angle, in a manner independent of explosion energy and circumburst density. We cover centroid motion, image shape and scale invarance in this work and present further discussion on how centroid motion and light curve slope measurements can be combined to constrain jet orientation and opening angle elsewhere \citep{Ryan:2023}.

The paper is organized as follows. In section \ref{section:scale_invariance}, we present generalized flux equations and some examples of how these can be applied. In section \ref{section:Deep_Newtonian}, we extend the scale invariance to the late time \emph{Deep Newtonian} emission regime where the assumption that the entire emitting electron population is relativistic no longer holds \citep{Granot:2006, SironiGiannios:2013}. The motion of the centroid and image shape are placed in the context of scale-invariance in section \ref{section:centroid}. We discuss our results in section \ref{section:discussion} and conclude in section \ref{section:conclusions}.

\section{Scale-invariant afterglow flux expressions}
\label{section:scale_invariance}

Canonically, afterglow models take the aforementioned parameters $E_0$, $\rho_0$ and $\theta_0$ to describe the initial conditions of the jet, along with a parametrization of the synchrotron emission using $p$ (as mentioned, the power-law slope of the non-thermal shock-accelerated electron population; typically $p \sim 2.2$), $\epsilon_e$ (the fraction of post-shock internal energy in the non-thermal electrons; typically $\epsilon_e \sim 10^{-1}$), $\epsilon_B$ (fraction of energy in the magnetic field; typically $\epsilon_B \sim 10^{-2}$ or lower) and $\xi_N$ (fraction of potentially available electrons participating in the non-thermal power-law distribution; typically $\xi_N \equiv 1$ to break a degeneracy in the model, \citealt{EichlerWaxman:2005}). 

As shown in \cite{vanEerten:2012scale-invariance}, the afterglow flux along any (intermediate) asymptote of the sychrotron spectrum can be rescaled in response to a change in $E_0$ and/or $\rho_0$. In that work, $E_0$ was taken to be the initial isotropic-equivalent energy of a top-hat jet (identical in value along angle within a top-hat jet), and $\rho_0$ the mass density of the circumburst medium. We generalize the former to indicate the isotropic-equivalent density specifically along the jet axis, allowing for arbitrary dimensionless functions $f(\theta)$ describing the energy profile $E(\theta)$ of the jet as a function of angle, such that $E(\theta) \equiv E_0 \times f(\theta)$. For example, the jet structure functions used in \cite{Ryan:2020} (i.e., implemented in \texttt{afterglowpy}) are given by
\begin{align}
    E(\theta) &= E_0 \times \exp\left(-\frac{1}{2}\left(\frac{\theta}{\thC}\right)^2\right) & &\text{Gaussian,}\label{eq:GaussianJet} \\
    E(\theta) &= E_0 \times \left(1 + \frac{1}{b}\left(\frac{\theta}{\thC}\right)^2\right)^{-b/2} & &\text{Power-Law.}\label{eq:PowerlawJet}
\end{align}
Here $\theta_c$ denotes the jet core angle as a measure of the characteristic width of the outflow and $b$ denotes the power-law index of the energy profile outside the core of a power-law jet profile. For top-hat jets, we can define $\thC \equiv \theta_0$ if needed to streamline notation.

The density profile can be generalized to describe an arbitrary radial power-law distribution of mass with density obeying $\rho (r) \equiv A r^{-k}$, where $k$ the power-law index of the distribution. For a homogeneous medium $k=0$ and $\rho_0$ is equivalent to $A$. For a non-zero value of $k$, such as $k = 2$ (modelling an environment shaped by a stellar wind) $A$ no longer has dimension of mass density but has $A = $ [g] [cm]$^{k-3}$.

The SRHD conservation laws can be written in terms of dimensionless coordinates $\mathcal{A}$, $\mathcal{B}$, $\theta$, $\phi$, where
\begin{equation}
\mathcal{A} \equiv \frac{r}{c t_e}, \quad 
\mathcal{B} \equiv \frac{E_0 t_e^2}{A r^{5-k}},
\label{eq:scaling_AB}
\end{equation}
instead of spherical polar coordinates $r$, $\theta$, $\phi$ and time $t_e$ in the burster frame (as opposed to observed time $t$). Here, $c$ denotes the speed of light. The dimensionless coordinates coordinates are manifestly invariant under any transformation
\begin{equation}
E_0' = \kappa E_0, \quad A' = \lambda A, \quad t_e' = \left( \frac{\kappa}{\lambda} \right)^{\frac{1}{3-k}} t_e, \quad r' = \left( \frac{\kappa}{\lambda} \right)^{\frac{1}{3-k}} r,
\label{eq:dynamical_scaling}
\end{equation}
as follows from the dimensions of $E_0$ and $A$ (scale-invariance expressions for general $k$ values were first presented in \citealt{vanEerten:2018}). When carried over to synchrotron spectra, the observed time $t$ follows the same scaling
\begin{equation}
t' = \left( \frac{\kappa}{\lambda} \right)^{\frac{1}{3-k}} t,
\label{eq:observer_time_scaling}
\end{equation}
which informs how the fluxes in different spectral regimes and the overall characteristic properties $F_{\textrm{peak}}$, $\nu_a$, $\nu_m$ and $\nu_c$ of the synchrotron spectrum can be rescaled to maintain an invariant outcome within an asymptotic spectral regime. 

Different spectral features will scale with different factors, given that they are computed from a synchrotron emission process that involves different combinations of dimension-carrying variables (including $c$, electron-mass $m_e$, proton mass $m_p$ and the Thomson cross section $\sigma_T$) for different spectral regimes. However, once this scaling is established, we know it has to apply throughout the evolution of the blast wave for any observation in the same spectral regime. We can illustrate this for an afterglow jet expanding into a homogeneous medium and a synchrotron flux $F_G$ observed at a frequency $\nu_{obs}$ for which $\nu_m < \nu_{obs} < \nu_c$ (as has so far been the case for GRB 170817A). If the dynamics are dictated by the self-similar ultra-relativistic limit from \cite{Blandford:1976}, this monochromatic flux $F_G$ obeys
\begin{equation}
F_{G,BM} \propto E_0^{\frac{p+3}{4}} \rho_0^{\frac{1}{2}} t^{\frac{3\left( 1-p\right)}{4}} \nu^{\frac{1-p}{2}}.
\end{equation}
According to Equations \ref{eq:dynamical_scaling} and \ref{eq:observer_time_scaling}, we therefore have a flux $F'_G (t')$ obeying
\begin{equation}
F'_G (t') = \kappa \lambda^{\frac{1+p}{4}} F_G(t),
\label{eq:F_G_example}
\end{equation}
once a rescaling in energy and density is matched with a rescaling of observed time. If Equation \ref{eq:F_G_example} is to be reproduced for a light curve of arbitrary slope, $F_G \propto t^\alpha$, we must have that
\begin{equation}
F'_G (t') \propto \kappa^{1-\frac{\alpha}{3-k}} \lambda^{\frac{1+p}{4}+\frac{\alpha}{3-k}} \left[ \left( \frac{\kappa}{\lambda} \right)^\frac{1}{3-k} t \right]^\alpha,
\end{equation}
and therefore that
\begin{equation}
F_G \propto E_0^{1-\frac{\alpha}{3}} \rho_0^{\frac{1+p}{4}+\frac{\alpha}{3}}  t^{\alpha} \nu^{\frac{1-p}{2}}.
\label{eq:F_G_full_example}
\end{equation}
As a quick cross-check, we can substitute the known temporal slope $\alpha = \frac{21-15p}{10}$ of the late non-relativistic Sedov-Taylor regime and find that
\begin{equation}
F_{G,ST} \propto E_0^{\frac{5p+3}{10}} \rho_0^{\frac{19-5p}{20}} t^{\frac{21-15p}{10}} \nu^{\frac{1-p}{2}},
\end{equation}
as expected. A more novel demonstration of the implications of Eq. \ref{eq:F_G_full_example} than the ST slope would be substituting $p = 2.17$ and rising slope $\alpha \approx 0.9$ (as inferred across frequencies for GRB 170817A before its turnover, see e.g. \citealt{Troja:2017, Troja:2018, Kasliwal:2017, Ruan:2018, Dobie:2018, D'Avanzo:2018}), we get
\begin{equation}
F \propto E_0^{0.7} \rho_0^{1.1} t^{0.9} \nu^{-0.585},
\end{equation}
showing how at this stage the light curve flux depends on energy somewhat less than linearly and close to linearly on environment density. 

Note in particular that this expression includes a generalization to a structured jet model of as yet unspecified structure but characterized by $E_0$, going beyond the top-hat or spherical explosion assumptions from previous works on scale invariance. If a Gaussian jet structure is assumed, we can cross-check this result against the flux equations provided in the appendix of \cite{Ryan:2020}. Substituting $S_\Omega = 1$ (applicable to the rising phase of a structured jet seen off-axis) and $g \sim 8.2$ (expressing a ratio between jet orientation and Gaussian core width inferred from the rising light curve slope) in Equation B24 from \cite{Ryan:2020} indeed recovers identical scalings for energy and density.

In tables \ref{table:scalings} and \ref{table:scalings_characteristics}, we provide a more extensive list of flux scaling relations. Unlike earlier work, these expressions make the dependency on temporal slope $\alpha$ explicit. The subscripts $A,B,D-H$ mark the observed spectral regime (also indicated in the table is how these relate to cooling break $\nu_c$, synchrotron injection break $\nu_m$ and self-absorption break $\nu_a$). They follow the same convention as in \cite{vanEertenWijers:2009, vanEerten:2012scale-invariance}, and can be traced back to \cite{GranotSari:2002}. The expressions in the table can be checked against the aforementioned papers, as well as \cite{Leventis:2012}. The scaling expressions make explicit the terms absorbed in the dimensionless functions of Table 1 in \cite{vanEertenMacFadyen:2013}, assuming these to be power laws (that work also introduces expressions for the case of a stellar-wind environment). A comparison with the same table of that paper also confirms the scaling expressions for the redshift term $1+z$, where we have made the dependency on $\alpha$ similarly explicit in the current work (note however how part of the $\alpha$-dependent redshift scaling is grouped with observer frame time $t$ in the current table, in order to emphasize the underlying structure of the expressions).

The flux expressions from table \ref{table:scalings} can be derived from the expressions for the synchrotron characteristics from table \ref{table:scalings_characteristics}. Note that $F_{\mathrm{peak}}$ is defined to be the peak flux either at the cooling break $\nu_c$ (fast cooling, when $\nu_c < \nu_m$) or at $\nu_m$ (slow cooling, when $\nu_c > \nu_m$), but without accounting for self-absorption. So if the self-absorption break occurs at a higher frequency than the leftmost of the pair $\nu_m$ and $\nu_c$, the actual peak of the spectrum will occur at $\nu_a$ and at a value lower than $F_{\mathrm{peak}}$. The evolution of the self-absorption break is different if $\nu_a > \nu_m$ rather than $\nu_a < \nu_m$, and a subscript number has been added to $\nu_a$ in the table to indicate which case applies. Because flux regime $B$ is agnostic to the relative ordering of $\nu_a$ and $\nu_m$, it is possible to express $\nu_{a2}$ in terms of $\nu_{a1}$ and $\nu_m$, as also indicated in the table:
\begin{equation}
\nu_{a2} \propto \nu_m^{\frac{2+3p}{3(p+4)}} \nu_{a1}^{\frac{10}{3(p+4)}}.
\end{equation}
This follows from the aforementioned definition of $F_{\mathrm{peak}}$ and equating two expressions for $F_B(\nu)$:
\begin{equation}
F_{\mathrm{peak}} \left( \frac{\nu_{a1}}{\nu_m} \right)^{\frac{1}{3}} \left( \frac{\nu}{\nu_{a1}} \right)^2 = F_{\mathrm{peak}} \left( \frac{\nu_{a2}}{\nu_m} \right)^{\frac{1-p}{2}} \left( \frac{\nu_m}{\nu_{a2}} \right)^{\frac{5}{2}} \left( \frac{\nu}{\nu_m} \right)^2.
\end{equation}
We have omitted the case where $\nu_a > \nu_c$, at which point the spectrum becomes very sensitive to the precise treatment of electron cooling \citep{GranotSari:2002} and the details of the radial structure of the blast wave, but the same principles apply here as well.

\begin{table*}
    \centering
    \begin{tabular}{lllll}
    $F_\nu$  & Scalings & $\kappa$ & $\lambda$ & spectral regime\\
    \hline \hline

    $F_{A,\mathrm{ISM}}$ & $\left( 1+z\right)^{\frac{7}{2}} E_0^{\frac{2-\alpha}{3}} \rho_0^{\frac{-11+4\alpha}{12}} \epsilon_e^0 \epsilon_B^{-\frac{1}{4}} \xi_N^0 \left( \frac{t}{1+z} \right)^\alpha \nu^{\frac{5}{2}}$ & $\kappa^{\frac{2}{3}}$ & $\lambda^{\frac{-11}{12}}$ & $\nu_m < \nu < \nu_a < \nu_c$\\

    $F_{A,\mathrm{wind}}$ & $\left( 1+z\right)^{\frac{7}{2}} E_0^{\frac{5-2\alpha}{2}} A^{\frac{-11+4\alpha}{4}} \epsilon_e^0 \epsilon_B^{-\frac{1}{4}} \xi_N^0 \left( \frac{t}{1+z} \right)^\alpha \nu^{\frac{5}{2}}$ & $\kappa^{\frac{5}{2}}$ & $\lambda^{\frac{-11}{4}}$ & $F_{\mathrm{peak}} \left( \frac{\nu_{a2}}{\nu_m} \right)^{\frac{1-p}{2}} \left( \frac{\nu}{\nu_{a2}} \right)^{\frac{5}{2}}$ \\

    $F_{A,k}$ & $\left( 1+z\right)^{\frac{7}{2}} E_0^{\frac{8+k-4\alpha}{4(3-k)}} A^{\frac{-11+4\alpha}{4(3-k)}} \epsilon_e^0 \epsilon_B^{-\frac{1}{4}} \xi_N^0 \left( \frac{t}{1+z} \right)^\alpha \nu^{\frac{5}{2}}$ & $\kappa^{\frac{k+8}{4(3-k)}}$ & $\lambda^{\frac{-11}{4(3-k)}}$ & \\

    \hline
    $F_{B, \mathrm{ISM}}$ & $\left( 1 + z\right)^{3} E_0^{\frac{2-\alpha}{3}} \rho_0^{\frac{-2+\alpha}{3}} \epsilon_e^{1} \epsilon_B^0 \xi_N^{-1} \left( \frac{t}{1+z} \right)^{\alpha} \nu^2$ & $\kappa^{\frac{2}{3}}$& $\lambda^{-\frac{2}{3}}$ & $\nu < \nu_a, \nu_m, \nu_c$  \\

    $F_{B, \mathrm{wind}}$ & $\left( 1 + z\right)^{3} E_0^{2-\alpha} A^{-2+\alpha} \epsilon_e^{1} \epsilon_B^0 \xi_N^{-1} \left( \frac{t}{1+z} \right)^{\alpha} \nu^2$ & $\kappa^{2}$& $\lambda^{-2}$ & $F_{\mathrm{peak}} \left( \frac{\nu_{a1}}{\nu_m} \right)^{\frac{1}{3}} \left( \frac{\nu}{\nu_{a1}} \right)^2$  \\

    $F_{B, k}$ & $\left( 1 + z\right)^{3} E_0^{\frac{2-\alpha}{3-k}} A^{\frac{-2+\alpha}{3-k}} \epsilon_e^{1} \epsilon_B^0 \xi_N^{-1} \left( \frac{t}{1+z} \right)^{\alpha} \nu^2$ & $\kappa^{\frac{2}{3-k}}$& $\lambda^{\frac{-2}{3-k}}$ &  \\

    \hline
    $F_{D, \mathrm{ISM}}$ & $(1+z)^{\frac{4}{3}} E_0^{\frac{3-\alpha}{3}} \rho_0^{\frac{1+\alpha}{3}} \epse^{-\frac{2}{3}} \epsB^{\frac{1}{3}} \xiN^{\frac{5}{3}} \left( \frac{t}{1+z} \right)^{\alpha} \nu^{\frac{1}{3}}$ & $\kappa^1$ & $\lambda^{\frac{1}{3}}$ & $\nu_a <  \nu < \nu_m < \nu_c$\\
    
    $F_{D, \mathrm{wind}}$ & $(1+z)^{\frac{4}{3}} E_0^{\frac{1-3\alpha}{3}} A^{1+\alpha} \epse^{-\frac{2}{3}} \epsB^{\frac{1}{3}} \xiN^{\frac{5}{3}} \left( \frac{t}{1+z} \right)^{\alpha} \nu^{\frac{1}{3}}$ & $\kappa^{\frac{1}{3}}$ & $\lambda^{1}$ & $F_{\mathrm{peak}} \left( \frac{\nu}{\nu_m} \right)^{\frac{1}{3}}$\\

    $F_{D, k}$ & $\left( 1 + z\right)^{\frac{4}{3}} E_0^{\frac{9-4k-3\alpha}{3(3-k)}} A^{\frac{1+\alpha}{3-k}} \epsilon_e^{-\frac{2}{3}} \epsilon_B^{\frac{1}{3}} \xi_N^{\frac{5}{3}} \left( \frac{t}{1+z} \right)^{\alpha} \nu^{\frac{1}{3}}$ & $\kappa^{\frac{9-4k}{3(3-k)}}$& $\lambda^{\frac{1}{3-k}}$ &  \\
    
    \hline
    $F_{E, \mathrm{ISM}}$ & $\left(1+z\right)^{\frac{4}{3}} E_0^{\frac{11-3\alpha}{9}} \rho_0^{\frac{7+3\alpha}{9}} \epse^{0} \epsB^{1} \xiN^{1} \left( \frac{t}{1+z} \right)^{\alpha} \nu^{\frac{1}{3}}$ & $\kappa^{\frac{11}{9}}$ & $\lambda^{\frac{7}{9}}$ & $\nu_a < \nu < \nu_c < \nu_m$ \\

    $F_{E, \mathrm{wind}}$ & $\left( 1 + z\right)^{\frac{4}{3}} E_0^{-\frac{1+3\alpha}{3}} A^{\frac{7+3\alpha}{3}} \epsilon_e^{0} \epsilon_B^1 \xi_N^1 \left( \frac{t}{1+z} \right)^{\alpha} \nu^{\frac{1}{3}}$ & $\kappa^{-\frac{1}{3}}$ & $\lambda^{\frac{7}{3}}$ & $F_{\mathrm{peak}} \left(\frac{\nu}{\nu_c}\right)^{\frac{1}{3}}$ \\

    $F_{E,k}$ & $\left( 1 + z\right)^{\frac{4}{3}} E_0^{\frac{11-6k-3\alpha}{3(3-k)}} A^{\frac{7+3\alpha}{3(3-k)}} \epsilon_e^{0} \epsilon_B^1 \xi_N^1 \left( \frac{t}{1+z} \right)^{\alpha} \nu^{\frac{1}{3}}$ & $\kappa^{\frac{11-6k}{3(3-k)}}$ & $\lambda^{\frac{7}{3(3-k)}}$ &  \\

    \hline
    $F_{F, \mathrm{ISM}}$ & $\left(1+z \right)^{\frac{1}{2}} E_0^{\frac{2-\alpha}{3}} \rho_0^{\frac{1+4\alpha}{12}} \epse^{0} \epsB^{-\frac{1}{4}} \xiN^{1} \left( \frac{t}{1+z} \right)^{\alpha} \nu^{-\frac{1}{2}}$ & $\kappa^{\frac{2}{3}}$ & $\lambda^{\frac{1}{12}}$ & $\nu_a, \nu_c < \nu < \nu_m$ \\

    $F_{F, \mathrm{wind}}$ & $\left( 1 + z\right)^{\frac{1}{2}} E_0^{\frac{1-2\alpha}{2}} A^{\frac{1+4\alpha}{4}} \epsilon_e^0 \epsilon_B^{-\frac{1}{4}} \xi_N^1 \left( \frac{t}{1+z} \right)^\alpha \nu^{-\frac{1}{2}}$ & $\kappa^{\frac{1}{2}}$ & $\lambda^{\frac{1}{4}}$ & $F_{\mathrm{peak}} \left( \frac{\nu}{\nu_c} \right)^{-\frac{1}{2}}$ \\

    $F_{F,k}$ & $\left( 1 + z\right)^{\frac{1}{2}} E_0^{\frac{8-3k-4\alpha}{4(3-k)}} A^{\frac{1+4\alpha}{4(3-k)}} \epsilon_e^0 \epsilon_B^{-\frac{1}{4}} \xi_N^1 \left( \frac{t}{1+z} \right)^\alpha \nu^{-\frac{1}{2}}$ & $\kappa^{\frac{8-3k}{4(3-k)}}$ & $\lambda^{\frac{1}{4(3-k)}}$ &  \\

    \hline
    $F_{G, \mathrm{ISM}}$ & $\left( 1+z \right)^{\frac{3-p}{2}} E_0^{\frac{3-\alpha}{3}} \rho_0^{\frac{3+3p+4\alpha}{12}} \epse^{p-1} \epsB^{\frac{1+p}{4}} \xiN^{2-p} \left( \frac{t}{1+z} \right)^{\alpha} \nu^{\frac{1-p}{2}}$ & $\kappa^1$& $\lambda^{\frac{1+p}{4}}$& $\nu_a, \nu_m < \nu < \nu_c$ \\

    $F_{G, \mathrm{wind}}$ & $\left( 1+z \right)^{\frac{3-p}{2}} E_0^{\frac{1-p- 2 \alpha}{2}} A^{\frac{3+3p+4\alpha}{4}} \epsilon_e^{p-1} \epsilon_B^{\frac{p+1}{4}} \xi_N^{2-p} \left(\frac{t}{1+z} \right)^{\alpha} \nu^{\frac{1-p}{2}}$ & $\kappa^{\frac{1-p}{2}}$ & $\lambda^{\frac{3(p+1)}{4}}$ & $F_{\mathrm{peak}} \left( \frac{\nu}{\nu_m} \right)^{\frac{1-p}{2}}$\\

    $F_{G,k}$ & $\left( 1+z \right)^{\frac{3-p}{2}} E_0^{\frac{12-pk-5k - 4 \alpha}{4(3-k)}} A^{\frac{3+3p+4\alpha}{4(3-k)}} \epsilon_e^{p-1} \epsilon_B^{\frac{p+1}{4}} \xi_N^{2-p} \left(\frac{t}{1+z} \right)^{\alpha} \nu^{\frac{1-p}{2}}$ & $\kappa^{\frac{12-5k-pk}{4(3-k)}}$ & $\lambda^{\frac{3(p+1)}{4(3-k)}}$ &  \\
    
    \hline
    $F_{H,\mathrm{ISM}}$ & $\left(1+z\right)^{\frac{2-p}{2}} E_0^{\frac{2-\alpha}{3}} \rho_0^{\frac{3p-2+4\alpha}{12}} \epse^{p-1} \epsB^{\frac{p-2}{4}} \xiN^{2-p} \left( \frac{t}{1+z} \right)^{\alpha} \nu^{-\frac{p}{2}}$ & $\kappa^{\frac{2}{3}}$ & $\lambda^{\frac{3p-2}{12}}$ & $\nu_a, \nu_m, \nu_c < \nu$ \\

    $F_{H, \mathrm{wind}}$ & $\left(1+z\right)^{\frac{2-p}{2}} E_0^{\frac{2-p-2\alpha}{2}} A^{\frac{3p-2+4\alpha}{4}} \epse^{p-1} \epsB^{\frac{p-2}{4}} \xiN^{2-p} \left( \frac{t}{1+z} \right)^{\alpha} \nu^{-\frac{p}{2}}$ & $\kappa^{\frac{2-p}{2}}$ & $\lambda^{\frac{3p-2}{4}}$ & $F_{\mathrm{peak}} \left( \frac{\nu_c}{\nu_m} \right)^{\frac{1-p}{2}} \left( \frac{\nu}{\nu_c} \right)^{-\frac{p}{2}}$ \\

    $F_{H,k}$ & $\left( 1+z \right)^{\frac{2-p}{2}} E_0^{\frac{8-2k-pk-4\alpha}{4(3-k)}} A^{\frac{3p-2+4\alpha}{4(3-k)}} \epsilon_e^{p-1} \epsilon_B^{\frac{p-2}{4}} \xi_N^{2-p} \left( \frac{t}{1+z} \right)^\alpha \nu^{-\frac{p}{2}}$ & $\kappa^{\frac{8-2k-pk}{4(3-k)}}$ & $\lambda^{\frac{3p-2}{4(3-k)}}$ &  \\
    \hline
    \end{tabular}
    \caption{Flux scaling expressions for the various spectral regimes. $\kappa$ and $\lambda$ are the energy and density rescale factor as defined in the text. The information in the spectral regime column applies to all $k$-values, indicating the spectral regime and how these regimes relate to the characteristic frequencies of the synchrotron spectrum.}
    \label{table:scalings}
\end{table*}

\begin{table*}
    \centering
    \begin{tabular}{lllll}
    $F_\nu$ or $\nu$ & Scalings & $\kappa$ & $\lambda$ & spectral regime \\

    \hline
    \hline
    $F_{\textrm{peak},\mathrm{ISM}}$ & $\left( 1 + z\right)^{1} E_0^{\frac{3-\alpha}{3}} \rho_0^{\frac{3+2\alpha}{6}} \epsilon_e^{0} \epsilon_B^{\frac{1}{2}} \xi_N^1 \left( \frac{t}{1+z} \right)^\alpha$ & $\kappa^1$ & $\lambda^{\frac{1}{2}}$ & \\
    
    $F_{\textrm{peak},\mathrm{wind}}$ & $\left( 1 + z\right)^{1} E_0^{- \alpha} A^{\frac{3}{2} + \alpha} \epsilon_e^{0} \epsilon_B^{\frac{1}{2}} \xi_N^1 \left( \frac{t}{1+z} \right)^\alpha$ & $\kappa^0$ & $\lambda^{\frac{3}{2}}$ &  \\

    $F_{\textrm{peak},k}$ & $\left( 1 + z\right)^{1} E_0^{\frac{6-3k-2\alpha}{2(3-k)}} A^{\frac{3+2\alpha}{2(3-k)}} \epsilon_e^{0} \epsilon_B^{\frac{1}{2}} \xi_N^1 \left( \frac{t}{1+z} \right)^\alpha$  & $\kappa^{\frac{3(2-k)}{2(3-k)}}$ & $\lambda^{\frac{3}{2(3-k)}}$ & \\
    \hline
    
    $\nu_{a1,\mathrm{ISM}}$ & $\left( 1+z\right)^{-1} E_0^{\frac{3-5\alpha}{15}} \rho_0^{\frac{9+5\alpha}{15}} \epsilon_e^{-1} \epsilon_B^{\frac{1}{5}} \xi_N^{\frac{8}{5}} \left( \frac{t}{1+z} \right)^{\alpha}$ & $\kappa^{\frac{1}{5}}$ & $\lambda^{\frac{3}{5}}$ & $\nu_{a1} < \nu_m < \nu_c$ \\

    $\nu_{a1, \mathrm{wind}}$ & $\left( 1+z\right)^{-1} E_0^{-1-\alpha} A^{\frac{9+5\alpha}{5}} \epsilon_e^{-1} \epsilon_B^{\frac{1}{5}} \xi_N^{\frac{8}{5}} \left( \frac{t}{1+z} \right)^{\alpha}$ & $\kappa^{-1}$ & $\lambda^{\frac{9}{5}}$ &  \\

    $\nu_{a1,k}$ & $\left( 1+z\right)^{-1} E_0^{\frac{3-4k-5\alpha}{15-5k}} A^{\frac{9+5\alpha}{15-5k}} \epsilon_e^{-1} \epsilon_B^{\frac{1}{5}} \xi_N^{\frac{8}{5}} \left( \frac{t}{1+z} \right)^{\alpha}$ & $\kappa^{\frac{3-4k}{5(3-k)}}$ & $\lambda^{\frac{9}{5(3-k)}}$ &  \\
    \hline

    $\nu_{a2,\mathrm{ISM}}$ & $\left( 1+z\right)^{-1} E_0^{\frac{2}{3(4+p)}-\frac{\alpha}{3}} \rho_0^{\frac{14+3p}{6(4+p)}+\frac{\alpha}{3}} \epsilon_e^{\frac{2(p-1)}{4+p}} \epsilon_B^{\frac{p+2}{2(4+p)}} \xi_N^{\frac{4-2p}{4+p}} \left( \frac{t}{1+z} \right)^{\alpha}$ & $\kappa^{\frac{2}{3(4+p)}}$ & $\lambda^{\frac{14+3p}{6(4+p)}}$ & $\nu_m < \nu_{a2} < \nu_c$ \\

    $\nu_{a2,\mathrm{wind}}$ & $\left( 1+z\right)^{-1} E_0^{-1-\alpha} A^{\frac{14+3p}{2(4+p)}+\alpha} \epsilon_e^{\frac{2(p-1)}{4+p}} \epsilon_B^{\frac{p+2}{2(4+p)}} \xi_N^{\frac{4-2p}{4+p}} \left( \frac{t}{1+z} \right)^{\alpha}$ & $\kappa^{-1}$ & $\lambda^{\frac{14+3p}{2(4+p)}}$ & $\nu_m^{\frac{2+3p}{3(p+4)}} \nu_{a1}^{\frac{10}{3(p+4)}}$\\

    $\nu_{a2,k}$ & $\left( 1+z\right)^{-1} E_0^{\frac{-(p+6)k-2\alpha p-8\alpha+4}{2(3-k)(4+p)}} A^{\frac{(2\alpha+3)p+8\alpha+14}{2(3-k)(4+p)}} \epsilon_e^{\frac{2(p-1)}{4+p}} \epsilon_B^{\frac{p+2}{2(4+p)}} \xi_N^{\frac{4-2p}{4+p}} \left( \frac{t}{1+z} \right)^{\alpha}$ & $\kappa^{\frac{4-(p+6)k}{2(3-k)(4+p)}}$ & $\lambda^{\frac{14+3p}{2(3-k)(4+p)}}$ &\\

    \hline
    $\nu_{m,\mathrm{ISM}}$ & $\left( 1 + z\right)^{-1} E_0^{-\frac{\alpha}{3}} \rho_0^{\frac{3+2\alpha}{6}} \epsilon_e^2 \epsilon_B^{\frac{1}{2}} \xi_N^{-2} \left( \frac{t}{1+z} \right)^\alpha$ & $\kappa^{0}$& $\lambda^{\frac{1}{2}}$& \\

    $\nu_{m,\mathrm{wind}}$ & $\left( 1 + z\right)^{-1} E_0^{-1-\alpha} A^{\frac{3+2\alpha}{2}} \epsilon_e^2 \epsilon_B^{\frac{1}{2}} \xi_N^{-2} \left( \frac{t}{1+z} \right)^\alpha$ & $\kappa^{-1}$& $\lambda^{\frac{3}{2}}$& \\

    $\nu_{m,k}$ & $\left( 1 + z\right)^{-1} E_0^{\frac{-(k+2\alpha)}{2(3-k)}} A^{\frac{3+2\alpha}{2(3-k)}} \epsilon_e^2 \epsilon_B^{\frac{1}{2}} \xi_N^{-2} \left( \frac{t}{1+z} \right)^\alpha$ & $\kappa^{\frac{-k}{2(3-k)}}$& $\lambda^{\frac{3}{2(3-k)}}$&  \\

    \hline
    $\nu_{c,\mathrm{ISM}}$ & $\left( 1 + z\right)^{-1} E_0^{-\frac{2+\alpha}{3}} \rho_0^{-\frac{5+2\alpha}{6}} \epsilon_e^{0} \epsilon_B^{-\frac{3}{2}} \xi_N^{0} \left( \frac{t}{1+z} \right)^\alpha$ & $\kappa^{-\frac{2}{3}}$ & $\lambda^{-\frac{5}{6}}$ & $\nu_c > \nu_a$  \\

    $\nu_{c, \mathrm{wind}}$ & $\left( 1 + z\right)^{-1} E_0^{1-\alpha} A^{\frac{-5+2\alpha}{2}} \epsilon_e^{0} \epsilon_B^{-\frac{3}{2}} \xi_N^{0} \left( \frac{t}{1+z} \right)^\alpha$ & $\kappa^{1}$ & $\lambda^{-\frac{5}{2}}$ &  \\

    $\nu_{c,k}$ & $\left( 1 + z\right)^{-1} E_0^{\frac{3k-4-2\alpha}{2(3-k)}} A^{\frac{-5+2\alpha}{2(3-k)}} \epsilon_e^{0} \epsilon_B^{-\frac{3}{2}} \xi_N^{0} \left(\frac{t}{1+z}\right)^\alpha$ & $\kappa^{\frac{3k-4}{2(3-k)}}$ & $\lambda^{\frac{-5}{2(3-k)}}$ & \\
    \hline
    \end{tabular}
\caption{Flux scaling expressions for characteristic features of the synchrotron spectrum. $F_{peak}$ is the peak flux in the absence of self-absorption. $\kappa$ and $\lambda$ are the energy and density rescale factor as defined in the text. The information in the spectral regime column applies to all $k$-values.}
\label{table:scalings_characteristics}
\end{table*}

\subsection{Some example applications}

\begin{figure*}
    \includegraphics[width=0.95\textwidth]{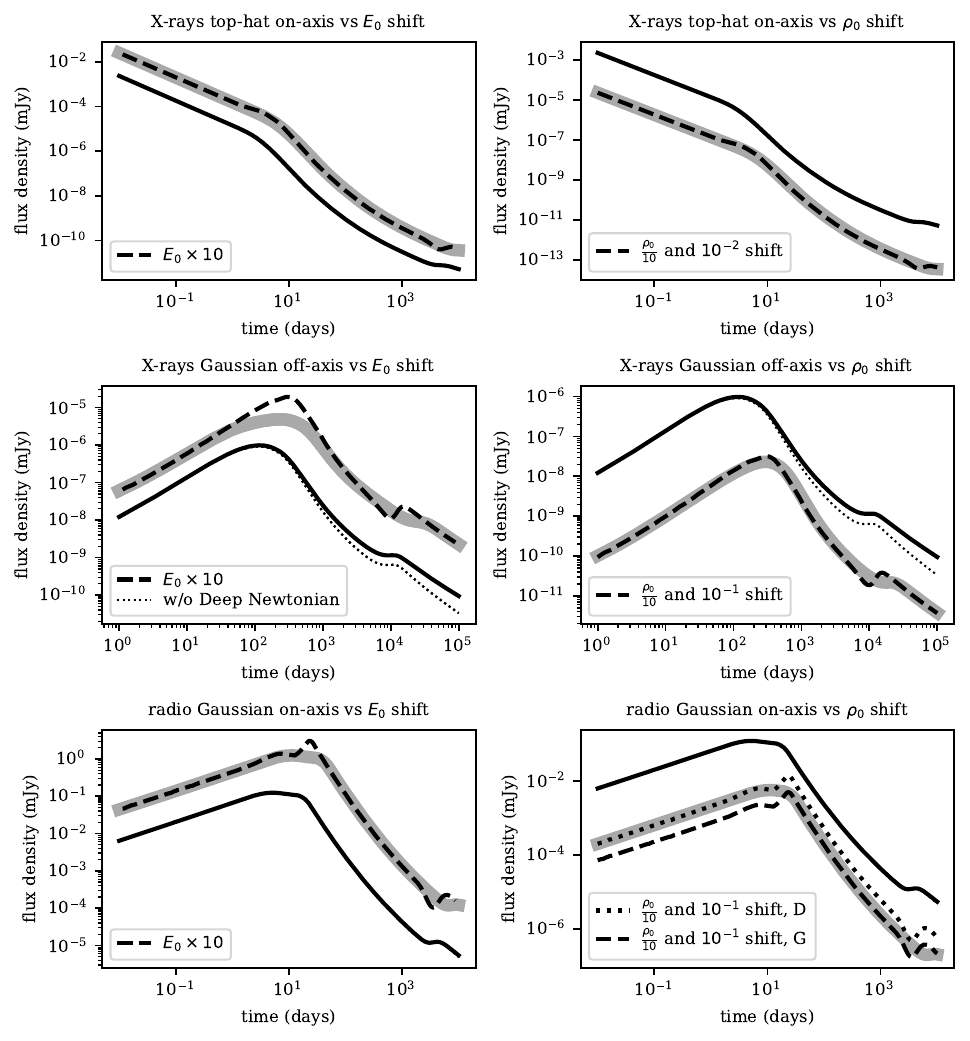}
    \caption{\label{fig:lightcurves} Example applications of rescalings. The solid curves and thick grey curves are produced using \texttt{afterglowpy}. The grey curves have their energy (left column) or density (right column) shifted as indicated in the figures. The dashed curves and (thick) dotted curve are produced by rescaling the solid curves based on local estimates of its power-law slope $\alpha$ and energy/density rescaling equations from table \ref{table:scalings} for the relevant spectral regime(s). Some grey and rescaled curves have been shifted further for clarity of presentation, as indicated in the figure legends. The thin dotted curves in the middle row are produced by not enabling the Deep Newtonian regime in \texttt{afterglowpy}. Top and bottom row solid curves illustrate a `typical' long GRB with model parameter values $E_0 = 10^{53}$ erg, $\rho_0 = m_p$ g cm$^{-3}$, $\epse = 3 \times 10^{-1}$, $\epsB = 10^{-3}$, $\xi_N = 1$, $p = 2.12$, luminosity distance $d_L = 5 \times 10^{28}$ cm and redshift $z=2$. The central row solid curves are illustrative of GRB 170817A, with parameter values $E_0 = 10^{54.35}$ erg, $\rho_0 = 10^{-1.75} m_p$ g cm$^{-3}$, $\epse = 10^{-3.42}$, $\epsB = 10^{-4.02}$, $\xi_N = 10^{-0.47}$, $p = 2.12$, $d_L = 40$ Mpc and $z=0$ (the median values from \citealt{Ryan:2023}). Top and middle row are computed at observer frequency $10^{18}$ Hz and the bottom row at $10^9$ Hz. The top row uses a top-hat jet with opening angle $\theta_C = 5.7^\circ$ (0.1 rad), the middle row a Gaussian jet with core width $\theta_C = 3.5^\circ$, truncation angle $\theta_W = 25^\circ$ and observed angle $\theta_{obs} = 20.8^\circ$, the bottom row a Gaussian jet with $\theta_C = 5.7^\circ$, $\theta_W =25^\circ$. Top and bottom jets are observed on-axis. Top row scalings assume spectral regime $H$, middle row scalings regime $G$ and bottom row both regimes $D$ and $G$ (the energy rescaling is identical for these). All environments are homogeneous in density.}
\end{figure*}

To the example of the rising slope $\alpha \sim 0.9$ of GRB 170817A more demonstrations of the implications of the scaling laws may be added. On-axis post jet break slopes have an $\alpha$ of about -2, which implies (for a homogeneous environment) a flux dependence of $F_G \propto E_{0}^{1.7} \rho^{0.13}$, telling us that this stage is strongly dependent on jet energy but barely dependent on circumburst density. We show a range of example applications in the panels of Figure \ref{fig:lightcurves}, spanning different spectral regimes and jet types. We emphasize that although we used \afterglowpy \ to produce the baseline light curve, the origin of the original light curve is irrelevant and we could for example have used an actual data set rather than a model-generated curve. 

We also stress that the procedure to produce the scaled curves (dashed curves and thick dotted curve in the figure) is completely agnostic as to the actual structure of the jet and works purely from an estimate of the local value of $\alpha$. The power law index $\alpha$ here has been estimated from a rudimentary comparison between adjacent synthetic data points, but the strengths (and limitations) of the rescalings as shown in the Figures are insensitive to this approach to $\alpha$.

The longer the stretch during which a light curve resembles an actual power law at fixed slope, the more accurate the rescaling manages to capture the fully recomputed curves (thick grey curves in the figure). This is expected, and we would be able to map the baseline curve on its recomputed counterpart \emph{exactly} by applying a scaling both in flux and time as described in \cite{vanEerten:2012scale-invariance}. 

However, even if the scaling approach from the current paper only involved an upward shift of the baseline curve rather than the diagonal one from \cite{vanEerten:2012scale-invariance}, the resulting light curves do end up with their characteristic features properly shifted in time. The peak positions of the curves in the bottom two rows of Figure \ref{fig:lightcurves} provide examples of this feature. The apparent horizontal shift in the original image is due to the dependence of the amount of vertical shift on the changing value of $\alpha$ for each horizontal start position.

Also seen at light curve peaks in particular is where the agnostic rescaling approaches is most limited. While the rescaled peak in the left panel of the middle row (where the baseline curve represents the X-ray afterglow of GRB 170817A) is indeed in the right place, it deviates noticeably from the fully recomputed curve. This is because $\alpha$ changes rapidly with time in the original curve around this point, which creates tension with the assumption of power law behaviour. A similar mismatch is apparent around the late time peaks associated with the appearance of the counterjet.

The scalings are the same between the `standard' approach to synchrotron emission and the Deep Newtonian limit that uses a different parametrization for synchrotron emission from a trans-relativistic electron population. The rescaled curves after the appearance of the counterjet (which will be in the Deep Newtonian regime in practice), can be seen to also match the fully recomputed curves. We will return to the Deep Newtonian regime in Section \ref{section:Deep_Newtonian}.

One does need to make an assumption about which spectral regime applies, however, which means an assumption about the relative order of the observed frequency $\nu_{\mathrm{obs}}$ and the synchrotron characteristic frequencies $\nu_a$, $\nu_m$ and $\nu_c$. Sometimes this does not matter. The bottom radio light curves in Figure \ref{fig:lightcurves} includes a transition from spectral regime $D$ to spectral regime $G$ (i.e. $\nu_m$ crosses the observer band). As the bottom left curve and table \ref{table:scalings} show, the energy scalings are identical between regimes D and F. On the other hand, the density scalings between the two \emph{do} differ, and this is shown in the bottom right panel. Regime D applies across the jet break (the first turnover), but a rescaling based on regime D ends up systematically overestimating its target value. A rescaling assuming regime G fails in the other direction on the left side of the transition. 

\subsubsection{Comparing the Swift XRT sample to model predictions}

\begin{figure}
    \includegraphics[width=\columnwidth]{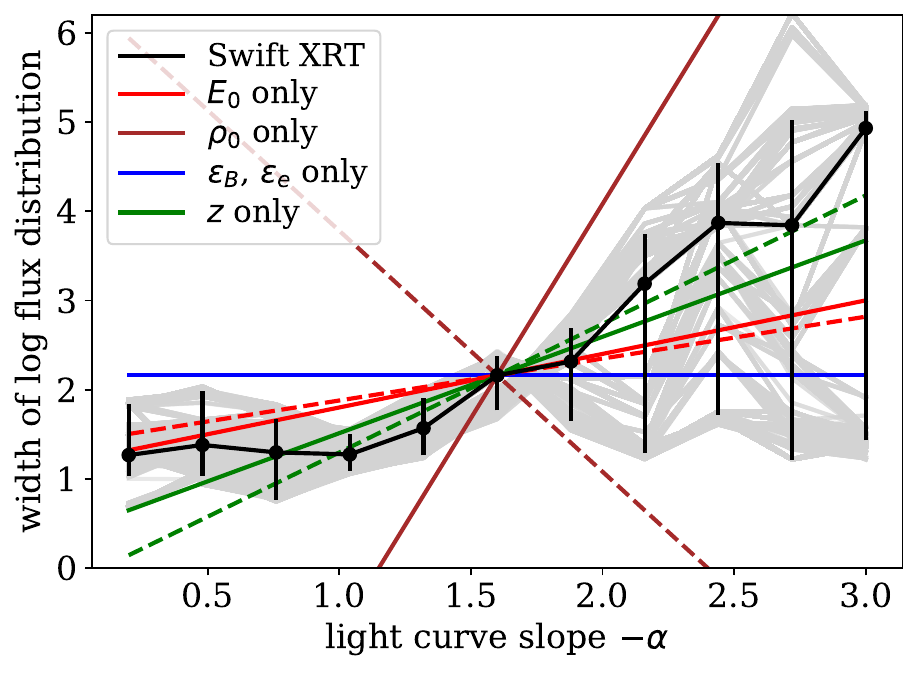}
    \caption{\label{fig:flux_distro}Width of the XRT sample flux distribution as a function of light curve slope. Black data points, black error bars and grey ranges are derived from the XRT sample as described in the main text. Coloured lines show the impact on the predicted distribution width when changing a single variable, centered on the middle data point. Solid lines assume spectral regime $H$ ($\nu > \nu_m, \nu_c$), dashed lines assume spectral regime $G$ ($\nu_m < \nu < \nu_c$). All coloured lines assume an ISM environment and a synchrotron slope value of $p=2.2$.}
\end{figure}

As demonstrated in the previous section, the local sensitivity of synchrotron light curves to a change in the value of an underlying physics parameter depends on its slope. It therefore follows that the range in flux values of a sample of light curves is dictated by the spread in parameter distributions realized in nature. To explore this, we have gathered information on the sample of afterglow X-ray light curves collected by the X-ray Telescope (XRT) onboard the Neil Gehrels Swift Observatory \citep{Gehrels:2004} during the period bracketed by GRB 041223 and GRB 231129A. We made use of the automatic summaries of XRT afterglows that are provided on-line at the Swift-XRT lightcurve repository\footnote{These data are available at \url{https://www.swift.ac.uk/xrt_curves/}, as part of the Swift-XRT GRB catalogue.} using the algorithm described in \cite{Evans:2007, Evans:2009}. For the sample of 1552 we downloaded the flux at 11 hours and the temporal slopes and break times as reported by the automated analysis. After discarding bursts for which no slope value and/or no 11 hour flux was reported, we were left with a sample of 1427 bursts.

This provided a basis for populating a distribution of flux values as function of light curve slope, where the flux values were computed from a reconstruction of the light curve using the 11 hr flux along with the slopes and break points. While this reduces each light curve to a very small set of temporal regimes and slopes (on average 2.25 slopes per light curve, for a total of 3213 slope values in the remaining sample), this has a few advantages over directly using the light curve data (photon counts or inferred flux values). First, we are guaranteed a reconstructed flux value at each observer time of interest within the duration of the afterglow, which makes it easier to compare fluxes across bursts. Second, the automated procedure removes features it identifies as flares before computing a connected power-law best fit. Third, the power-law reconstructions capture the trends of the light curves, rather than the fluctuations in the data due to measurement error or genuine short-term variability.

The grey and black lines and data points in Figure \ref{fig:flux_distro} show the range in fluxes across the sample against light curve slope $-\alpha$. These were obtained from the reconstructed light curves described above, sampled at 2000 random times between 10 and $10^6$ seconds (with draws evenly distributed in logarithmic space). For each time sampled, we order the flux draws within each of the 11 slope bins between $-\alpha = 0.2$ and $-\alpha = 3.0$ by magnitude and compare the values of the fluxes $F_{84}$ and $F_{16}$ that have respectively 16\% and 84\% of the values below them, to compute a measure of the width $W$ of the distribution:
\begin{equation}
W(\alpha, t) \equiv \log_{10} F_{84} - \log_{10} F_{16}. 
\end{equation}
Between them, $F_{84}$ and $F_{16}$ capture $66\%$ of the flux values within the bin (for some observer time), which is the range of fluxes covered by a one-sigma deviation from the median if the flux values were following a Gaussian distribution in log space.

All $W$ values are plotted in grey in Figure \ref{fig:flux_distro}, using straight lines to connect the values at the different $\alpha$ values for each given time. Plotted in black are, for each $\alpha$ bin, the median of the population of $W(\alpha, t)$ values across sample times, again connected with lines. The vertical black error bars cover $32\%$ above and below the median.

If the observed range in flux values were fully determined by some physics parameter that affects the light curve independent of its local slope, the trend in Figure \ref{fig:flux_distro} would have been flat. Examples of such parameters are the synchrotron parameters $\epsilon_e$ and $\epsilon_B$, shown in the figure with a blue line centered on the middle $\alpha$ bin (note how the exponents in the scalings with respect to $\epsilon_e$ and $\epsilon_B$ in table \ref{table:scalings} do not contain factors of $\alpha$ referring to light curve slope). The other colored lines in Figure \ref{fig:flux_distro}, however, show the impact of physics parameters that do depend on local slope, with dashed lines assuming synchrotron regime $G$ (observations below the cooling break) and solid lines assuming synchrotron regime $H$ (observations above the cooling break).

The actual trend displayed by the data is a superposition of the impacts of the different physics parameters, with the relative weight of each parameter being set by the range of values for this parameter manifested in nature. A few things stand out in the figure, even if the error bars increase rapidly with steeper light curve decay slopes. The spread looks to be larger for steeper values of $a$, running counter to at least the dependence on density in regime $G$ in an ISM environment (i.e. the dashed curve brown trending down. The ISM dependence on density in regime $H$ trends steeply upwards. This indicates that the range in densities represented in the light curves is only minor or that assuming a homogeneous environment is not sufficiently accurate. A complication is that the slope of the model dependencies for density and energy in turn generally depend on electron power-law distribution slope $p$.

\subsubsection{Shifting with redshift}

\begin{figure}
    \includegraphics[width=\columnwidth]{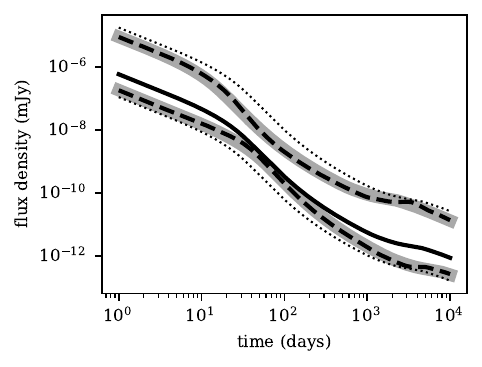}
    \caption{\label{fig:redshifts} A demonstration of rescaling with redshift for a synthetic X-ray ($\nu = 10^{18}$ Hz) light curve observed in regime $H$ above the cooling break, computed using \afterglowpy. Solid curve shows a Gaussian jet light curve at $z = 2$ ($d_L = 15.8$ Gpc), with model parameters $\theta_{obs} = 10^\circ$, $\theta_C = 5^\circ$, $\theta_W = 25^\circ$, $E_0 = 10^{53}$ erg, ISM environment $\rho_0 = 1 \ m_p$, $p = 2.2$, $\epsilon_e = 10^{-1}$, $\epsilon_B = 10^{-3}$, $\xi_N = 1$. Upper and lower thick grey curves show the same light curve recomputed at redshift $z=0.5$ ($d_L = 2.86$ Gpc) and $z = 4$ ($d_L = 36.6$ Gpc) respectively. The dashed curves are produced by rescaling the $z=2$ curve using the regime $H$ redshift scaling from table \ref{table:scalings} and accounting for the difference in luminosity distance. Dotted curves only account for the luminosity distance difference.}
\end{figure}

A final application of rescaling is shown in Figure \ref{fig:redshifts}, demonstrating a rescaling of a light curve across redshift values (and corresponding luminosity distances), akin to a generalized K correction. For this demonstration we have used a Gaussian structured jet observed at an angle. To convert redshifts to luminosity distances, we have assumed a standard cosmology with $H_0 = 69.6$, $\Omega_M = 0.286$, $\Omega_\Lambda = 0.714$. The dotted curves in the Figure only include an accounting for the difference in luminosity distance, which emphasizes the impact of the redshift rescaling shown with dashed curves. In the Figure, we rescale a light curve computed at $z=2$ to both $z = 0.5$ and $z=4$.

\section{The Deep Newtonian case}
\label{section:Deep_Newtonian}

So far we have followed the default approach to modelling of the accelerated electron population as a power law in energy, $n_e \sim \gamma_e^{-p}$ between a lower cut-off Lorentz factor $\gamma_m$ and an upper cut-off Lorentz factor whose actual value has negligible impact on the energy total if the power-law slope $p$ is sufficiently steep (i.e., larger than 2). Ignoring the upper cut-off Lorentz factor, $\gamma_m$ can be expressed as:
\begin{equation}
\gamma_m = \frac{p-2}{p-1} \frac{\epse e }{\xiN n m_e c^2} = \frac{\bar{\epsilon}_e}{\xiN} \frac{e}{n m_e c^2}.
\label{eq:gamma_m}
\end{equation}
Here, $e$ is the energy density of the shocked plasma and $n$ is the total electron number density of the plasma in the frame of the fluid (and equal to proton number density, to ensure charge neutrality). Electron density $n$ and internal energy density $e$ are dictated by the jump conditions across the shock front. The terms with $p$ are often absorbed into the energy fraction, as indicated with $\bar{\epsilon}_e$ (see e.g. \citealt{GranotSari:2002}). Doing so has the advantage that the formalism can now be extended to cover values of $p$ smaller than 2, where the upper cut-off can no longer be ignored for the purpose of determining the total energy of the electron population even if occurring at high electron energy. This comes at the cost of not being able to interpret $\bar{\epsilon}_e$ as a fraction of energy.   

There comes a point in the late-stage evolution where the prescription from Equation \ref{eq:gamma_m} inevitably breaks down, potentially while still in the strong shock regime. Using the transrelativistic equation-of-state (EOS) from \citealt{MignonePlewaBodo:2005}, we find that setting $\gamma_m \sim 1$ implies\footnote{Using $\gamma_m -1$ on the LHS of Equation \ref{eq:gamma_m} will avoid a situation where $\gamma_m < 1$, but will still eventually lead to a power law distribution in energies extending to unrealistically low value.}:

\begin{align}
    &1 = \gamma_m = \frac{\bar{\epsilon}_e}{\xiN} \frac{m_p}{m_e} \left( \gamma - 1 \right)  \\
    &\implies \beta \sim \sqrt{2 \frac{\xiN}{\bar{\epsilon}_e} \frac{m_e}{m_p}} \sim 0.33 \left(\frac{\bar{\epsilon}_e}{0.01}\right)^{-1/2} \left(\frac{\xiN}{1.0}\right)^{1/2}\ . \nonumber
    \label{eq:gamma_m_limit}
\end{align}

Here $\beta$ is the fluid velocity in units of $c$. If $\xiN \approx 1$, then the standard afterglow prescription predicts \emph{non}-relativistic shocked electrons once the blast wave decelerates to $\beta \sim 0.33$.  The lower $\xiN$, the longer this breakdown is delayed (see also \citealt{vanEerten:2010} for an example of an approach where $\xiN$ is lowered over time).

\cite{Granot:2006} and \cite{SironiGiannios:2013} offer a means to extend the afterglow synchrotron model to allow for regimes where the lower cut-off $\gamma_m$ approaches unity (structured jets like GRB 170817A render this issue more relevant due to the decreasing energy in the wings of the jet reaching the non-relativistic stage earlier than the tip; for further discussion see \citealt{Ryan:2023}). The core idea from \cite{SironiGiannios:2013} is to acknowledge that a power-law distribution in electron Lorentz factor $\gamma_e$ (as a proxy for electron energy $\gamma_e m_e c^2$) represents the relativistic limit of what more generally should be a power law in momentum instead. This is implemented in an approximate manner by fixing $\gamma_m$ at 1 when it reaches this value according to Equation \eqref{eq:gamma_m}, along with a change in the proportionality of the emission and absorption coefficients for synchrotron emission. Instead of setting these proportional to the number density of emitters (and to the usual factor depending on synchrotron slope $p$), $\propto (p - 1) \xiN n$, they are taken to follow $\propto (p -2) \epse e / \left( \gamma_m m_e c^2 \right)$ throughout the entire evolution of the blast wave. As can be shown from the shock-jump conditions across a strong relativistic shock, the generalized proportionality factor reduces to the standard proportionality in the relativistic limit.

The newly introduced behaviour is labeled the \emph{Deep Newtonian} limit \citep{SironiGiannios:2013, HuangCheng:2003} (DN), and leads generally to a shallower decay of the late-time light curves. Because the synchrotron emission from the blast wave is now modelled using a different parametrisation than before, the expressions provided in table \ref{table:scalings} for the scalings of $\epsilon_e$, $\epsilon_B$ and $\xi_N$ are no longer applicable. The Deep Newtonian limit can be viewed as effectively introducing additional synchrotron regimes, and we summarize these in tables \ref{table:scalings_DN} and \ref{table:scalings_DN_characteristics}. Note that the scalings with energy and density (the central topic of this work), remain unchanged relative to the ST regime. In these tables, therefore it is the scaling with respect to $\xi_N$ and $\epsilon_e$ that is novel relative to tables \ref{table:scalings} and \ref{table:scalings_characteristics} and that has not been presented in explicit form in the literature before.

In the tables we include the expected temporal slopes values $\alpha$ for the different spectral regimes and density profiles. By the time the jet emission segues into the DN regime, beaming and collimation no longer play a role in shaping the emission and jet structure is therefore no longer expected to influence the slope $\alpha$. In this asymptotic regime, which in practice will take a long time to achieve, the dynamics are driven by the Sedov-Taylor-Von Neumann limit. The cooling break is assumed to be unaffected by the Deep Newtonian transition, as it describes the point in electron-energy space where injection and radiative losses from (fast-moving) electrons are in balance. Between regime $D$ and regime $G$ the connection is across the Deep Newtonian peak frequency $\nu_{pk} \propto t^{-3/5}$ (obtained when fixing $\gamma_m \sim 2$). In the Deep Newtonian regime, $\xiN$ is still a measure of the total fraction of accelerated electrons, but not all of these electrons are radiating.

\begin{table*}
    \centering
    \begin{tabular}{lllll}
    $F_\nu$ & Scalings & $\kappa$ & $\lambda$ & DN $\alpha$ \\
    \hline \hline

    $F_{A,\mathrm{ISM},\mathrm{DN}}$ & $\left( 1+z\right)^{\frac{7}{2}} E_0^{\frac{2-\alpha}{3}} \rho_0^{\frac{-11+4\alpha}{12}} \epsilon_e^0 \epsilon_B^{-\frac{1}{4}} \xi_N^0 \left( \frac{t}{(1+z)} \right)^{\alpha} \nu^{\frac{5}{2}}$ & $\kappa^{\frac{2}{3}}$ & $\lambda^{-\frac{11}{12}}$ & $\frac{11}{10}$ \\

    $F_{A,\mathrm{wind},\mathrm{DN}}$ & $\left( 1+z\right)^{\frac{7}{2}} E_0^{\frac{5-2\alpha}{2}} A^{\frac{-11+4\alpha}{4}} \epsilon_e^0 \epsilon_B^{-\frac{1}{4}} \xi_N^0 \left( \frac{t}{(1+z)} \right)^{\alpha}  \nu^{\frac{5}{2}}$ & $\kappa^{\frac{5}{2}}$ & $\lambda^{-\frac{11}{4}}$ & $\frac{11}{6}$ \\

    $F_{A,k,\mathrm{DN}}$ & $\left( 1+z\right)^{\frac{7}{2}} E_0^{\frac{8+k-4\alpha}{4(3-k)}} A^{\frac{-11+4\alpha}{4(3-k)}} \epsilon_e^0 \epsilon_B^{-\frac{1}{4}} \xi_N^0 \left( \frac{t}{(1+z)} \right)^{\alpha}  \nu^{\frac{5}{2}}$ & $\kappa^{\frac{8+k}{4(3-k)}}$ & $\lambda^{\frac{-11}{4(3-k)}}$ & $\frac{11}{2(5-k)}$ \\
    \hline

    $F_{B, \mathrm{ISM},\mathrm{DN}}$ & $\left( 1+z \right)^3 E_0^{\frac{2-\alpha}{3}} \rho_0^{\frac{-2+\alpha}{3}} \epsilon_e^0 \epsilon_B^0 \xi_N^0 \left( \frac{t}{1+z} \right)^{\alpha} \nu^2$ & $\kappa^{\frac{2}{3}}$ & $\lambda^{-\frac{2}{3}}$ & $\frac{4}{5}$ \\

    $F_{B, \mathrm{wind},\mathrm{DN}}$ & $\left( 1+z \right)^3 E_0^{2-\alpha} \rho_0^{-2+\alpha} \epsilon_e^0 \epsilon_B^0 \xi_N^0 \left( \frac{t}{1+z} \right)^{\alpha} \nu^2$ & $\kappa^{2}$ & $\lambda^{-2}$ & $\frac{4}{3}$ \\

    $F_{B,k,\mathrm{DN}}$ & $\left( 1+z \right)^3 E_0^{\frac{2-\alpha}{3-k}} A^{\frac{-2+\alpha}{3-k}} \epsilon_e^0 \epsilon_B^0 \xi_N^0 \left( \frac{t}{1+z} \right)^{\alpha} \nu^2$ & $\kappa^{\frac{2}{3-k}}$ & $\lambda^{\frac{-2}{3-k}}$ & $\frac{4}{5-k}$ \\
    \hline

    $F_{D, \mathrm{ISM},\mathrm{DN}}$ & $\left( 1 + z\right)^{\frac{4}{3}} E_0^{\frac{3-\alpha}{3}} \rho_0^{\frac{1+\alpha}{3}} \epse^{1} \epsB^{\frac{1}{3}} \xiN^{0} \left( \frac{t}{1+z} \right)^\alpha \nu^{\frac{1}{3}}$ & $\kappa^1$ & $\lambda^{\frac{1}{3}}$ & $-\frac{2}{5}$ \\

    $F_{D,\mathrm{wind},\mathrm{DN}}$ & $\left( 1 + z\right)^{\frac{4}{3}} E_0^{\frac{1-3\alpha}{3}} A^{1+\alpha} \epse^{1} \epsB^{\frac{1}{3}} \xiN^{0}\left( \frac{t}{1+z} \right)^\alpha \nu^{\frac{1}{3}}$ & $\kappa^{\frac{1}{3}}$ & $\lambda^{1}$ & $-\frac{2}{3}$ \\

    $F_{D,k,\mathrm{DN}}$ & $\left( 1 + z\right)^{\frac{4}{3}}  E_0^{\frac{9-4k-3\alpha}{3(3-k)}} A^{\frac{1+\alpha}{3-k}} \epse^{1} \epsB^{\frac{1}{3}} \xiN^{0}\left( \frac{t}{1+z} \right)^\alpha \nu^{\frac{1}{3}}$ & $\kappa^{\frac{9-4k}{3(3-k)}}$& $\lambda^{\frac{1}{3-k}}$ & $-\frac{2}{5-k}$ \\

    \hline

    $F_{G,\mathrm{ISM},\mathrm{DN}}$ & $\left( 1+z \right)^{\frac{3-p}{2}} E_0^{\frac{3-\alpha}{3}} \rho_0^{\frac{3+3p+4\alpha}{12}} \epse^{1} \epsB^{\frac{1+p}{4}} \xiN^{0} \left( \frac{t}{1+z} \right)^{\alpha} \nu^{\frac{1-p}{2}}$ & $\kappa^1$& $\lambda^{\frac{1+p}{4}}$ & $-\frac{3(1+p)}{10}$ \\

    $F_{G,\mathrm{wind}, \mathrm{DN}}$ & $\left( 1+z \right)^{\frac{3-p}{2}} E_0^{\frac{1-p- 2 \alpha}{2}} A^{\frac{3+3p+4\alpha}{4}} \epsilon_e^{1} \epsilon_B^{\frac{p+1}{4}} \xi_N^{0} \left(\frac{t}{1+z} \right)^{\alpha} \nu^{\frac{1-p}{2}}$ & $\kappa^{\frac{1-p}{2}}$ & $\lambda^{\frac{3(p+1)}{4}}$ & $-\frac{1+p}{2}$ \\

    $F_{G,k, \mathrm{DN}}$ & $\left( 1+z \right)^{\frac{3-p}{2}} E_0^{\frac{12-pk-5k - 4 \alpha}{4(3-k)}} A^{\frac{3+3p+4\alpha}{4(3-k)}} \epsilon_e^{1} \epsilon_B^{\frac{p+1}{4}} \xi_N^0 \left(\frac{t}{1+z} \right)^{\alpha} \nu^{\frac{1-p}{2}}$ & $\kappa^{\frac{12-5k-pk}{4(3-k)}}$ & $\lambda^{\frac{3(p+1)}{4(3-k)}}$ & $-\frac{3(1+p)}{2(5-k)}$ \\

    \hline

    $F_{H,\mathrm{ISM},\mathrm{DN}}$ & $\left(1+z\right)^{\frac{2-p}{2}} E_0^{\frac{2-\alpha}{3}} \rho_0^{\frac{3p-2+4\alpha}{12}} \epse^{1} \epsB^{\frac{p-2}{4}} \xiN^{0} \left( \frac{t}{1+z} \right)^{\alpha} \nu^{-\frac{p}{2}}$ & $\kappa^{\frac{2}{3}}$ & $\lambda^{\frac{3p-2}{12}}$ & $-\frac{4+3p}{10}$ \\

    $F_{H,\mathrm{wind},\mathrm{DN}}$ & $\left(1+z\right)^{\frac{2-p}{2}} E_0^{\frac{2-p-2\alpha}{2}} A^{\frac{3p-2+4\alpha}{4}} \epse^{1} \epsB^{\frac{p-2}{4}} \xiN^{0} \left( \frac{t}{1+z} \right)^{\alpha} \nu^{-\frac{p}{2}}$ & $\kappa^{\frac{2-p}{2}}$ & $\lambda^{\frac{3p-2}{4}}$ & $-\frac{p}{2}$ \\

    $F_{H,k,\mathrm{DN}}$ & $\left( 1+z \right)^{\frac{2-p}{2}} E_0^{\frac{8-2k-pk-4\alpha}{4(3-k)}} A^{\frac{3p-2+4\alpha}{4(3-k)}} \epsilon_e^{1} \epsilon_B^{\frac{p-2}{4}} \xi_N^{0} \left( \frac{t}{1+z} \right)^\alpha \nu^{-\frac{p}{2}}$ & $\kappa^{\frac{8-2k-pk}{4(3-k)}}$ & $\lambda^{\frac{3p-2}{4(3-k)}}$ & $\frac{-4-3p+2k}{2(5-k)}$ \\    

    \hline
    \end{tabular}
    \caption{Flux scaling expressions for the various spectral regimes, Deep Newtonian case. Spectral regimes are the same as in table \ref{table:scalings}, but with $\nu_{pk}$ taking the place of $\nu_m$. The final column shows the temporal slope of the flux in the Deep Newtonian limit.}
    \label{table:scalings_DN}
\end{table*}

\begin{table*}
    \centering
    \begin{tabular}{lllll}
    $F_\nu$ or $\nu$ & Scalings & $\kappa$ & $\lambda$ & DN $\alpha$ \\
    \hline \hline
    $F_{\textrm{peak},\mathrm{ISM},\mathrm{DN}}$ & $\left( 1 + z\right)^1 E_0^{\frac{3-\alpha}{3}} \rho_0^{\frac{3+2\alpha}{6}} \epsilon_e^1 \epsilon_B^{\frac{1}{2}} \xi_N^0 \left( \frac{t}{1+z} \right)^\alpha$ & $\kappa^{1}$ & $\lambda^{\frac{1}{3}}$ & $-\frac{3}{5}$ \\

    $F_{\textrm{peak},\mathrm{wind},\mathrm{DN}}$ & $\left( 1 + z\right)^1 E_0^{-\alpha} A^{\frac{3+2\alpha}{2}} \epsilon_e^1 \epsilon_B^{\frac{1}{2}} \xi_N^0 \left( \frac{t}{1+z} \right)^\alpha$ & $\kappa^{0}$ & $\lambda^{\frac{3}{2}}$ & $-1$ \\

    $F_{\textrm{peak},k,\mathrm{DN}}$ & $\left( 1 + z\right)^1 E_0^{\frac{6-3k-2\alpha}{2(3-k)}} A^{\frac{3+2\alpha}{2(3-k)}} \epsilon_e^1 \epsilon_B^{\frac{1}{2}} \xi_N^0 \left( \frac{t}{1+z} \right)^\alpha$ & $\kappa^{\frac{6-3k}{2(3-k)}}$ & $\lambda^{\frac{3}{2(3-k)}}$ & $-\frac{3}{5-k}$ \\
    
    \hline

    $\nu_{a1,\mathrm{ISM}, \mathrm{DN}}$ & $\left( 1+z\right)^{-1} E_0^{\frac{3-5\alpha}{15}} \rho_0^{\frac{9+5\alpha}{15}} \epsilon_e^{\frac{3}{5}} \epsilon_B^{\frac{1}{5}} \xi_N^{0} \left( \frac{t}{1+z} \right)^{\alpha}$ & $\kappa^{\frac{1}{5}}$ & $\lambda^{\frac{3}{5}}$ & $-\frac{18}{25}$ \\

    $\nu_{a1,\mathrm{wind},\mathrm{DN}}$ & $\left( 1+z\right)^{-1} E_0^{-1-\alpha} A^{\frac{9+5\alpha}{5}} \epsilon_e^{\frac{3}{5}} \epsilon_B^{\frac{1}{5}} \xi_N^{0} \left( \frac{t}{1+z} \right)^{\alpha}$ & $\kappa^{-1}$ & $\lambda^{\frac{9}{5}}$ & $-\frac{6}{5}$ \\

    $\nu_{a1,k,\mathrm{DN}}$ & $\left( 1+z\right)^{-1} E_0^{\frac{3-4k-5\alpha}{15-5k}} A^{\frac{9+5\alpha}{15-5k}} \epsilon_e^{\frac{3}{5}} \epsilon_B^{\frac{1}{5}} \xi_N^{0} \left( \frac{t}{1+z} \right)^{\alpha}$ & $\kappa^{\frac{3-4k}{5(3-k)}}$ & $\lambda^{\frac{9}{5(3-k)}}$ & $-\frac{18}{5(5-k)}$ \\
    \hline

    $\nu_{a2,\mathrm{ISM},\mathrm{DN}}$ & $\left( 1+z\right)^{-1} E_0^{\frac{2}{3(4+p)}-\frac{\alpha}{3}} \rho_0^{\frac{14+3p}{6(4+p)}+\frac{\alpha}{3}} \epsilon_e^{\frac{2}{4+p}} \epsilon_B^{\frac{p+2}{2(4+p)}} \xi_N^{0} \left( \frac{t}{1+z} \right)^{\alpha}$ & $\kappa^{\frac{2}{3(4+p)}}$ & $\lambda^{\frac{14+3p}{6(4+p)}}$ & $-\frac{14+3p}{5(4+p)}$ \\

    $\nu_{a2,\mathrm{wind},\mathrm{DN}}$ & $\left( 1+z\right)^{-1} E_0^{-1-\alpha} A^{\frac{14+3p}{2(4+p)}+\alpha} \epsilon_e^{\frac{2}{4+p}} \epsilon_B^{\frac{p+2}{2(4+p)}} \xi_N^{0} \left( \frac{t}{1+z} \right)^{\alpha}$ & $\kappa^{-1}$ & $\lambda^{\frac{14+3p}{2(4+p)}}$ & $-\frac{14+3p}{3(4+p)}$\\

    $\nu_{a2,k,\mathrm{DN}}$ & $\left( 1+z\right)^{-1} E_0^{\frac{-(p+6)k-2\alpha p-8\alpha+4}{2(3-k)(4+p)}} A^{\frac{(2\alpha+3)p+8\alpha+14}{2(3-k)(4+p)}} \epsilon_e^{\frac{2}{4+p}} \epsilon_B^{\frac{p+2}{2(4+p)}} \xi_N^{0} \left( \frac{t}{1+z} \right)^{\alpha}$ & $\kappa^{\frac{4-(p+6)k}{2(3-k)(4+p)}}$ & $\lambda^{\frac{14+3p}{2(3-k)(4+p)}}$ & $-\frac{14+3p}{(4+p)(5-k)}$ \\
    
    \hline
    $\nu_{pk,\mathrm{ISM},\mathrm{DN}}$ & $\left( 1 + z\right)^{-1} E_0^{\frac{-\alpha}{3}} \rho_0^{\frac{3+2\alpha}{6}} \epsilon_e^0 \epsilon_B^{\frac{1}{2}} \xi_N^0 \left( \frac{t}{1+z} \right)^{\alpha}$ & $\kappa^0$ & $\lambda^{\frac{1}{2}}$ & $-\frac{3}{5}$ \\
    
    $\nu_{pk,\mathrm{wind},\mathrm{DN}}$ & $\left( 1 + z\right)^{-1} E_0^{-1-\alpha} A^{\frac{3+2\alpha}{2}} \epsilon_e^0 \epsilon_B^{\frac{1}{2}} \xi_N^0 \left( \frac{t}{1+z} \right)^{\alpha}$ & $\kappa^{-1}$ & $\lambda^{\frac{3}{2}}$ & $-1$ \\
    
    $\nu_{pk,k,\mathrm{DN}}$ & $\left( 1 + z\right)^{-1} E_0^{\frac{-k-2\alpha}{2(3-k)}} A^{\frac{3+2\alpha}{2(3-k)}} \epsilon_e^0 \epsilon_B^{\frac{1}{2}} \xi_N^0 \left( \frac{t}{1+z} \right)^{\alpha}$ & $\kappa^{\frac{-k}{2(3-k)}}$ & $\lambda^{\frac{3}{2(3-k)}}$ & $-\frac{3}{5-k}$ \\
    \hline
    $\nu_{c,\mathrm{ISM},\mathrm{DN}}$ & $\left( 1 + z\right)^{-1} E_0^{-\frac{2+\alpha}{3}} \rho_0^{-\frac{5+2\alpha}{6}} \epsilon_e^{0} \epsilon_B^{-\frac{3}{2}} \xi_N^{0} \left( \frac{t}{1+z} \right)^\alpha$ & $\kappa^{-\frac{2}{3}}$ & $\lambda^{-\frac{5}{6}}$ & $-\frac{1}{5}$ \\

    $\nu_{c,\mathrm{wind},\mathrm{DN}}$ & $\left( 1 + z\right)^{-1} E_0^{1-\alpha} A^{\frac{-5+2\alpha}{2}} \epsilon_e^{0} \epsilon_B^{-\frac{3}{2}} \xi_N^{0} \left( \frac{t}{1+z} \right)^\alpha$ & $\kappa^{1}$ & $\lambda^{-\frac{5}{2}}$ & $1$ \\

    $\nu_{c,k,\mathrm{DN}}$ & $\left( 1 + z\right)^{-1} E_0^{\frac{3k-4-2\alpha}{2(3-k)}} A^{\frac{-5+2\alpha}{2(3-k)}} \epsilon_e^{0} \epsilon_B^{-\frac{3}{2}} \xi_N^{0} \left(\frac{t}{1+z}\right)^\alpha$ & $\kappa^{\frac{3k-4}{2(3-k)}}$ & $\lambda^{\frac{-5}{2(3-k)}}$ & $\frac{-1+2k}{5-k}$\\
    \hline    
    \end{tabular}
    \caption{Flux scaling expressions for characteristic features of the synchrotron spectrum, for the Deep Newtonian case. $F_{peak}$ is the peak flux in the absence of self-absorption. $\kappa$ and $\lambda$ are the energy and density rescale factor as defined in the text. The final column shows the time dependence of the characteristic quantity in the Deep Newtonian regime.    }
    \label{table:scalings_DN_characteristics}

\end{table*}

\section{Scale-invariance, sky image and centroid motion}
\label{section:centroid}

The VLBI observations of GRB 170817A point to another quantity of interest associated with GRB afterglows to which the rescalings from Equations \ref{eq:dynamical_scaling} must be applicable: the angular offset $\xcen$ (relative to its origin at $t = 0$) of the centroid of the image on the sky produced by the afterglow. Indeed, Equations \ref{eq:dynamical_scaling} apply to the entire dynamics and thus to the whole image on the sky produced by the jet, but as a practical matter, the centroid offset $\xcen$ in particular is an observable that allows for very tight constraints on the combination of jet orientation and core width. 

In brief, the observed flux $F_\nu$ can be expressed as an integral of the specific intensity $I_\nu$ over the observer's sky: $F_\nu = \int I_\nu d\Omega$.  The centroid of the image $\xcen$ is just the intensity weighted average of the angular position $\tilde{x}$, that is $\xcen = \int \tilde{x} I_\nu d\Omega / F_\nu$.  Given the angular diameter distance $d_A$ to the source, the proper displacement of the centroid is then $x_c = d_A \xcen$ (see \citealt{Ryan:2023}, for further discussion and for details on how the centroid is computed in \afterglowpy).  

The effective image sizes $\tilde{\sigma}_x$ (in the direction parallel to the jet propagation) and $\tilde{\sigma}_y$ (transverse to the jet propagation) can be computed from the second moments of the intensity: $\tilde{\sigma}_x^2 = \int (\tilde{x}-\xcen)^2 I_\nu d\Omega / F_\nu$ and similar for $\tilde{\sigma}_y^2$.  The effective circular size is $\tilde{\sigma}^2 = \tilde{\sigma}_x^2 + \tilde{\sigma}_y^2$.

Observations of the centroid motion and effective image size have been used to constrain models of GRB afterglows. A first breakthrough event was GRB 030329, for which \cite{Taylor:2004} were able to demonstrate apparent superluminal expansion using analytical modelling. More recently, \cite{Mooley:2018} have used VLBI data for GRB 170817A to argue that its apparent superluminal motion is consistent with a compact source, based on a combination of hydrodynamic and point source modelling of the outflow. In a follow-up, \cite{Hotokezaka:2019} confirm and complement this modelling with a semi-analytical approach. \cite{Gill:2019} point out that the apparent velocity is dimensionless and therefore invariant under the rescaling of energy and circumburst density also discussed in this paper, although measurements of the same apparent velocity for differing values of $E$ and $\rho$ corresponds to different observer times.

The centroid position and image size are measures of linear distances in the GRB blast wave.  Identical to the scaling of the other times and distances in the system $t_e$, $r$ and $t$, we have
\begin{align}
    \xcen' &= \left( \frac{\kappa}{\lambda} \right)^{\frac{1}{3-k}} \xcen ,&
    \tilde{\sigma}_x' &= \left( \frac{\kappa}{\lambda} \right)^{\frac{1}{3-k}} \tilde{\sigma}_x, &
    \tilde{\sigma}_y' &= \left( \frac{\kappa}{\lambda} \right)^{\frac{1}{3-k}} \tilde{\sigma}_y .
\end{align}

Much like the light curve, interpretation of a measure of the centroid is complicated by energy, density, distance, jet orientation and the overall time evolution of the blast as dictated by its structure and dynamics. We therefore propose potential observables that at least utilize the scalings of the system in a way that makes for a more universal measure: the average centroid and expansion velocities from launch to the time of the jet break $t_{\mathrm{jb}}$, and the ellipticity $\tilde{\sigma}_x / \tilde{\sigma}_y$ at the same time. The jet break is a key characteristic feature of the light curve that follows the time scaling from Equation \ref{eq:observer_time_scaling}. In the case of a jet observed strongly off-axis (the most relevant case as far as VLBI observations of GW counterparts are concerned), this break refers to the peak of the light curve, a generic and unambiguous feature expected from jets regardless of their structure. The jet break therefore represents either an advantageous time for a VLBI measurement at peak brightness of the source, or the last time before the flux begins to diminish more rapidly in an observation closer to on-axis (\citealt{Granot:2018}, by comparison, make a simular point about invariance using the transition time to non-relativistic flow as a normalizing factor).  These particular observables are dimensionless combinations of $\xcen$, $\tilde{\sigma}_x$, $\tilde{\sigma}_y$, and $t_{\mathrm{jb}}$ and so their dependence on $\kappa$ and $\lambda$ cancel out.

\begin{figure}
    \includegraphics[width=\columnwidth]{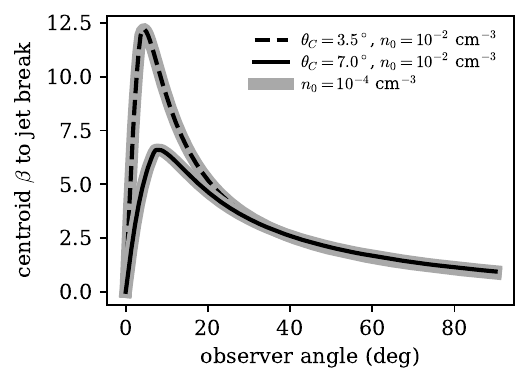}
    \caption{\label{fig:centroid} Average centroid velocity between launch and jet break time, for two Gaussian jets of different core width. For both jets the curves are computed for different circumburst number densities $n_0 = \rho_0 / m_p$, to illustrate that the outcome is invariant under density and/or energy rescaling.}
\end{figure}

Positioning the centroid at $\xcen = 0$ at the start time $t=0$ of the explosion and the jet propagation to be in the $\tilde{x}$- direction, we have:
\begin{align}
    \beta_{\mathrm{avg}} &\equiv x_c(t_{\mathrm{jb}}) / c t_{\mathrm{jb}} & &\text{average centroid velocity} , \\
    \beta_{\sigma, \mathrm{avg}} &\equiv \sigma(t_{\mathrm{jb}}) / c t_{\mathrm{jb}} & &\text{average expansion velocity} , \\
    e_{\mathrm{jb}} &\equiv \tilde{\sigma}_x(t_{\mathrm{jb}}) / \tilde{\sigma}_y(t_{\mathrm{jb}}) & &\text{ellipticity at jet break} .
\end{align} 
Being scale-invariant ($\beta_{\mathrm{avg}}' = \beta_{\mathrm{avg}}$, $\beta_{\sigma, \mathrm{avg}}' = \beta_{\sigma, \mathrm{avg}}$, $e_{\mathrm{jb}}' = e_{\mathrm{jb}}$) and independent of distance, there is an appealing universality to the distribution of the these quantities with jet orientation. We provide examples of the centroid velocity in Figure \ref{fig:centroid}. To construct the figure, the jet break has been computed using Equation 34 from \cite{Ryan:2020}. The curves in the figure are invariant under changes in density (or energy) by construction. As can be seen from figure, for a given jet the average centroid velocity peaks when $\theta_{\mathrm{obs}} \approx \theta_C$. This is in line with the expectation for a point source rather than a jet. For $\theta_{\mathrm{obs}} \gg \theta_C$, the distinction between different $\theta_C$ values becomes negligible.

It can be demonstrated that the shape of the curve does not depend strongly on the structure of the jet at high observer angles. However, having demonstrated scale-invariance here, we defer further discussion of this aspect to \cite{Ryan:2023}, where we also elaborate on how centroid measurements can be combined with flux observations.

\begin{figure}
    \includegraphics[width=\columnwidth]{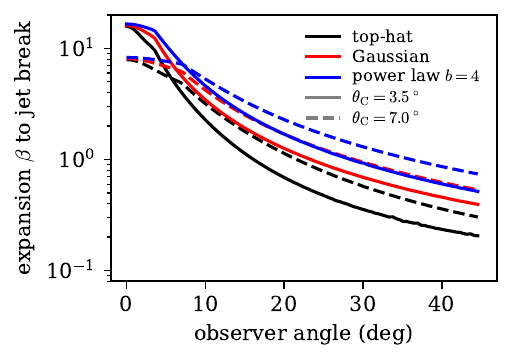}
    \caption{\label{fig:size} Average expansion velocity between launch and jet break time, for a top-hat jet (black), Gaussian jet (red), and $b=4$ power law jet (blue), each with opening angles of $3.5^\circ$ (solid) and $7^\circ$ (dashed).}
\end{figure}

\begin{figure}
    \includegraphics[width=\columnwidth]{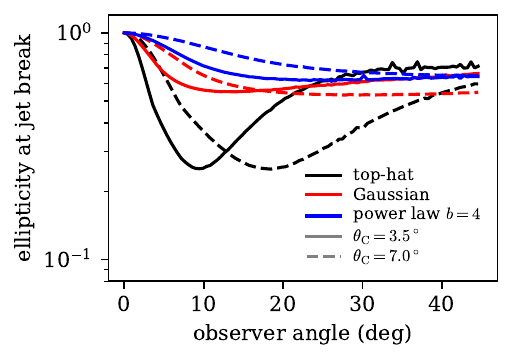}
    \caption{\label{fig:ellipticity} Image ellipticity at the jet break time, for a top-hat jet (black), Gaussian jet (red), and $b=4$ power law jet (blue), each with opening angles of $3.5^\circ$ (solid) and $7^\circ$ (dashed).}
\end{figure}

Figures \ref{fig:size} and \ref{fig:ellipticity} show $\beta_{\sigma, \mathrm{avg}}$ and $e_{\mathrm{jb}}$ as a function of observer inclination for several different jet structures.  As in Figure \ref{fig:centroid}, these curves are independent of $E_0$ and $\rho_0$.  For the image size, on-axis viewers see an apparent expansion speed inversely proportional to the size of the core of the jet: $\beta_{\sigma, \mathrm{avg}}(\theta_{\rm obs} = 0) \sim \gamma(t_{\rm jb}) \sim 1/\thC$.  At larger viewing angles the image size at the jet break falls in a structure-dependent way, generally faster for narrower structures.  Since our jet models are axisymmetric, on-axis all jets appear circular: $e_{\rm jb} = 1$. As the viewing angle increases, narrower jets show stronger ellipticity, shortened in the direction of propagation, to a minimum of $\sim 0.22$ for top-hat jets.  At inclinations greater than $\sim 20^\circ$ many models show a roughly constant ellipticity between 0.6 and 0.8.

\section{Discussion}
\label{section:discussion}

The scaling methods presented in this work have all been demonstrated from \afterglowpy \ and a shell model for a blast wave during its deceleration phase. Nevertheless, rescaling with energy and/or density would work identically during phases of energy injection in the afterglow, as is demonstrated in \cite{vanEerten:2014}, where the flux expressions are given for a self-similar Blandford-McKee solution including strong reverse shock and ultra-relativistic wind of ongoing injection. It remains true though that, like jet structure, energy injection adds further model parameters that increase the diversity of light curves that can be produced. Any additional parameters do remain subject to the same scaling relations as the ones showed in this work, given that these are based on dimensional analysis. A light curve plateau end time $t_p$ (which might be identified with the cessation time of energy injection), will inevitably scale according to $t_p' = \left( \kappa / \lambda \right)^{1/(3-k)} t_p$, et cetera.

All the scalings presented here are relative and inferring an actual value for $\rho_0$, $A$ or $E_0$ still requires assuming an underlying model to calibrate against. Some stages of light curve evolution are more suitable to this exercise than others. Late-stage afterglow calorimetry, for example, relies on the assumption of a transition to non-relativistic flow, when the shape of the emitting volume no longer matters (see e.g. \citealt{Frail:2005}). 

Nevertheless, even if releative, the scalings can be used to explore the extent to which the diversity in a sample of light curves can be attributed to an intrinsic range of density, energy or redshift values in the burst population. We have used the Swift XRT sample to illustrate the concept. A more in-depth study along these lines is beyond the scope of the current work, but example avenues for exploration could include a direct fit of multiple power-law contributions from the different physics parameters, weighted by the width of the underlying distributions. The translation from XRT data to connected power-law light curves has been treated in a basic manner, again to provide an illustration of the principle rather than an in-depth study. Finally, observational biases and selection effects might result in the \emph{observed} width of a parameter distribution being not completely identical across light curve slope bins (steep slopes predominantly cover the light curves at later, fainter stages, for example).

A practical issue when accounting for the full set ($\epse, \epsB, \xiN$ as well as $E_0$ and $A$) of microphysical model parameters is that there exists a degeneracy between them. The exact same synchrotron spectrum is reproduced for any value of $q$ in the set $q \xiN$, $q \epse$, $q \epsB$, $E_0 / q$ and $A / q$ \citep{EichlerWaxman:2005}.  The degeneracy between model parameters stays intact when the Deep Newtonian limit is included in the model:  $\gamma_m$ remains unchanged under shifts in $q$ and the new emission/absorption proportionalities respond in an identical manner to the old. These invariances can be confirmed from Tables \ref{table:scalings}, \ref{table:scalings_characteristics}, \ref{table:scalings_DN} and \ref{table:scalings_DN_characteristics} by applying when taking $\xi_N \to q \xi_N$, $\epsilon_e \to q \epsilon_e$, $\epsilon_B \to q \epsilon_B$, $E_0 \to q^{-1} E_0$ and $A \to q^{-1} A$ and finding the flux expressions unaffected. A fit, for example, with fixed $\xi_N \equiv 1$ therefore does not measure $E_0$ itself but merely a lower limit.

An obvious caveat applies to our discussion of the added information from VLBI measurements from section \ref{section:centroid}, which is that such observations are going to remain exceedingly rare in practice. Short GRB and GW-counterpart GRB 170817A remains so far unique in providing us with a measurement of the centroid motion \citep{Mooley:2018, Ghirlanda:2019, Mooley:2022}, whereas observations of the expansion of the ejecta are up to now limited to the long GRB cases GRB 030329 \citep{Taylor:2004} and GRB 221009A \citep{Giarratana:2023}. In principle, one can extend the predictions from studies of the rates of GRBs (e.g. \citealt{Beniamini:2019, Salafia:2023}) to predictions for centroid motion and image detectability through VLBI, although this is beyond the scope of the current work. Most likely off-axis detections of GRBs will be for observer angles within the wings of the jet.

\section{Conclusions}
\label{section:conclusions}

In this work, we build upon scaling equations presented in a series of papers starting with \cite{vanEerten:2012boxfit} and \cite{vanEerten:2012scale-invariance}. We show that once a synchrotron spectral regime is assumed (perhaps inferred from an observed spectral slope or photon index) and a density profile is assumed (i.e., a value for $k$ in $\rho \propto r^{-k}$ is chosen), the proportionality of a light curve with respect to energy and density is fully determined. This holds regardless of the underlying structure of the jet (Gaussian, power-law, top-hat or other), its dynamical stage of evolution (relativistic, non-relativistic or in between, collimated or not) or whether a synthetic model-generated light curve or an actual data set is taken as starting point. Ultimately, the scaling relations are expressions of basic dimensional analysis, albeit obscured by the details of the physics of the synchrotron power-law spectrum. A series of flux scaling relations are presented in Tables \ref{table:scalings}, \ref{table:scalings_characteristics}, \ref{table:scalings_DN} and \ref{table:scalings_DN_characteristics}, with the latter two tables covering the Deep Newtonian regime. The scalings with energy and density are identical between both regimes.

Sky images of afterglows can be rescaled in the same manner, and we propose the average velocity $\beta_{\mathrm{avg}}$ of the centroid between jet launch and jet break (the peak of a highly off-axis event such as GRB 170817A) as a robust and scale-invariant observable. Generalized curves can be constructed of $\beta_{\mathrm{avg}}$ versus observer angle $\theta_{\mathrm{obs}}$ that are independent of all model parameters except for the core angle $\theta_C$ of a Gaussian structured jet or an equivalent that sets the scale of lateral energy distribution in a jet of arbitrary structure.

Flux scaling relations can be used to assess the sensitivity of observed light curves to changes in the scales of their underlying physics and to map the diversity of light curves, without the limiting assumption of a particular jet model. We have demonstrated this for the Swift XRT sample, finding a broadening of the observed flux range for steeper slopes. This is consistent with an underlying range of redshift and explosion energy values, while the evidence from this feature for a broad range in underlying environment densities (assuming an ISM circumburst medium) is less strong and opposed to the expected trend if the XRT sample is assumed to represent light curves observed below the cooling break.

The average centroid velocity can be used to constrain the opening angle of the jet and can be a powerful means to break model degeneracies in multi-messenger observations (see \citealt{Ryan:2023} for further details). The size and ellipticity of the image change rapidly in a structure-dependent way, faster for narrower structures, as the viewing angle increases from on-axis observations.

Although the results in this paper are all presented in terms of GRB afterglows, we note that our flux equations and conclusions are generic to any synchrotron transient characterized by a release of an energy $E_0$ in an external medium described by $A$ and $k$. As such, this work is applicable to e.g. supernova remnants, kilonova afterglows and soft gamma-repeater flares as well.

\section*{Acknowledgements}

The authors thank Nora Troja and Luigi Piro for helpful discussion and HJvE thanks for their hospitality the Perimeter Institute for Theoretical Physics at Waterloo, Canada, where a large part of this work was completed. Research at Perimeter Institute is supported in part by the Government of Canada through the Department of Innovation, Science and Economic Development and by the Province of Ontario through the Ministry of Colleges and Universities. HJvE acknowledges support by the European Union horizon 2020 programme under the AHEAD2020 project (grant agreement number 871158) and by the STFC through grant ST/X001067/1.

\section*{Data Availability}

No new data were generated or analysed in support of this research.





\bibliographystyle{mnras}
\bibliography{general_case} 





\bsp	
\label{lastpage}
\end{document}